\documentclass[prl,aps,twocolumn,superscriptaddress]{revtex4-2}

\usepackage{bm,bbm,color,float,dcolumn,amssymb,graphicx,subfigure, mathtools, mhchem}

\usepackage[colorlinks=true]{hyperref}
\hypersetup{linkcolor=blue,citecolor=blue,urlcolor=blue}

\graphicspath{ {./figures/} }

\DeclarePairedDelimiter\bra{\langle}{\vert}
\DeclarePairedDelimiter\ket{\vert}{\rangle}
\DeclarePairedDelimiter\xpct{\langle}{\rangle}
\DeclarePairedDelimiterX\braket[2]{\langle}{\rangle}{#1 \delimsize\vert #2}

\definecolor{violet}{rgb}{0.56, 0, 1}

\begin{document}

\title{Quantum Mpemba effect from Stark localization}

\author{Nico Albert}
\affiliation{Institut f\"ur Theoretische Physik, Technische Universit\"at Dresden, 01062 Dresden, Germany}

\author{Masudul Haque}
\affiliation{Institut f\"ur Theoretische Physik, Technische Universit\"at Dresden, 01062 Dresden, Germany}

\author{Shovan Dutta}
\affiliation{Raman Research Institute, Bangalore 560080 India}

\date{\today}

\begin{abstract}
In classical systems, rugged potential energy landscapes provide a transparent mechanism for the celebrated Mpemba effect, in which hotter initial states cool down faster. 
This picture generally does not survive in quantum systems.  Here we show how to design a quantum energy landscape with local dissipation leading to an Mpemba effect with parametrically separated timescales.  
Our approach uses Stark localization to design an energy landscape and localized incoherent hopping as cooling mechanism. The hops are triggered by rare ``detection'' events whose rate grows with energy, allowing hotter states to cool faster and producing super-exponentially separated cooling rates for localized initial states. We further show that the effect is dramatically enhanced by collective hopping of bound pairs in the presence of attractive on-site interactions. These findings have clear experimental signatures accessible in present-day setups.
\end{abstract}

\maketitle

Understanding the pathways and timescales by which open quantum systems approach their steady states is a central problem in nonequilibrium quantum dynamics. Particularly intriguing are situations in which relaxation is strongly state dependent, leading to metastability \cite{Macieszczak2016_OpenMetastability, Minganti2018, Macieszczak2021_ClassicalMetastability, Rose2022, Zhang2025_metastability} and anomalous ordering of relaxation times. A prominent example, which has attracted considerable attention in recent years, is the Mpemba effect \cite{Bechhoefer2021, Teza2026, Ares_Calabrese_Murciano_NatRevPhys2025_Mpemba_review, Calabrese2026}, in which an initial state farther from steady state relaxes more quickly. Originally studied in the context of cooling in classical systems (e.g., colloids), it has been generalized to both closed \cite{Ares2023, Murciano2024, Rylands2024, Liu2024, Turkeshi2025, Bhore2025, Muller2026, Calabrese2026} and open \cite{Nava_Fabrizio_PRB2019_Mpemba, Carollo_Lesanovsky_PRL2021_Mpemba, Moroder2024, Ivander_Segal_PRE2023_Mpemba, Chatterjee_Hayakawa_PRL2023_Mpemba, Nava_Egger_PRL2024_Mpemba, Wang2024_Mpemba, Longhi_APLQ2024_Mpemba, Furtado_Santos_AnnPhys2025_Mpemba, Strachan2025, DiGiulio2025, ganguly2026measurement, ulvcakar2025conserved, Zhao2026, Ali2026, Wei2026} quantum systems and observed in ion traps \cite{Joshi2024, AharonyShapira2024, Zhang2025, xia2026observation}, NMR systems \cite{chatterjee2025_NMRMpemba} and superconducting circuits \cite{Xu2026}. Identifying transparent physical mechanisms underlying this effect is of fundamental interest \cite{Rylands2024, Summer2026, yamashika2026quantum} and may also enable new ways to control relaxation pathways and metastability in open quantum systems.

In classical systems, an intuitive and widely applicable mechanism for the thermal Mpemba effect is provided by rugged potential-energy landscapes \cite{Lu_Raz_pnas2017_Mpemba}, where a system close to a local minimum must climb uphill before cooling toward the global minimum. Such local minima act as metastable states whose escape rates depend strongly on the surrounding barriers, whereas higher-energy states can cool faster. While thermal Mpemba effects have also been identified in quantum systems \cite{li2025canonical, Furtado_Santos_AnnPhys2025_Mpemba, Chatterjee_Hayakawa_PRL2023_Mpemba, Wei2026}, this landscape picture lacks a quantum counterpart (except in semiclassical limits \cite{Nava_Fabrizio_PRB2019_Mpemba}).  In a quantum system,  energy eigenstates need not be localized on the potential landscape. Moreover, local dissipation does not ordinarily produce a simple, directed flow through the energy spectrum. Thus, the correspondence between potential minima, metastable states, and barrier-controlled cooling rates is generally lost.

Here we show how one can circumvent this obstacle by combining Stark localization with local incoherent hopping. Localization on a tilted lattice \cite{Hartmann2004} provides a direct correspondence between single-particle energy and position, allowing one to engineer {\it total-energy} landscapes by spatially varying the local tilt. Incoherent hopping provides a cooling mechanism by inducing transitions between these localized states. By restricting such hops to a neighborhood of the ground state, we generate an extensive hierarchy of long-lived states and a pronounced Mpemba effect, in which higher-energy states can cool (super)exponentially faster. This hierarchy of relaxation times is intrinsically quantum: it originates from the exponentially small tails of the localized eigenstates, and has no counterpart in the corresponding classical random walk, where transport is instead diffusive.

Remarkably, the effect survives in the presence of interactions, even though the one-to-one mapping between energy and position is lost. For two interacting bosons, beyond a critical interaction strength the relaxation pathway changes from individual to collective hopping. This strongly enhances the Mpemba effect for attractive interactions, while suppressing it for repulsive interactions.

Experiments have already realized interacting bosons on Stark ladders with variable tilt \cite{Simon2011, Meinert_Naegerl_PRL2014_BlochOscilExpt, Preiss_Greiner_Science2015_quantumwalks, Adler2024}, and there are concrete protocols to realize incoherent hopping \cite{Haga_Ueda_PRL2021_LiouvillianSkinEffect, Sharma2021, Garbe2024}. 
Thus, our results provide a transparent and experimentally accessible mechanism to engineer a tunable quantum Mpemba effect with a hierarchy of relaxation rates, and more broadly demonstrate how localization, dissipation, and interactions can be combined to design quantum energy landscapes and control relaxation pathways.

\begin{figure}
\resizebox{\columnwidth}{!}{\includegraphics{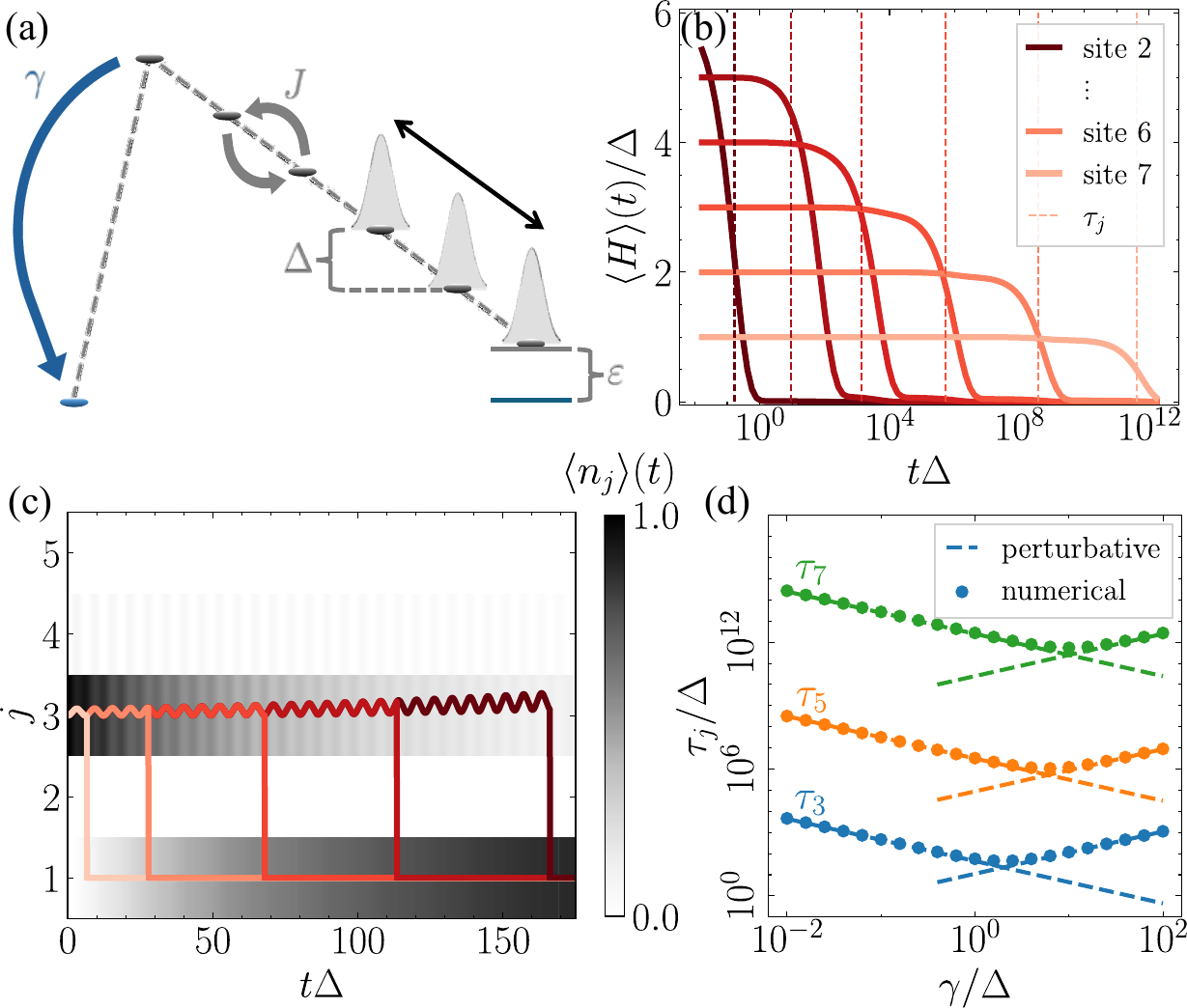}}
\caption{(a) Schematic of single-particle setup, illustrating rates $\gamma$, $J$ and energy gaps  $\Delta$, $\varepsilon$.  (b) Time evolution of the energy for initial states localized on different sites. High energy states close to site 2 relax faster than low energy states that are localized farther to the right.  The $\tau_j$ are relaxation times obtained from Eqs.\  \eqref{eq:relaxation_times_small}, \eqref{eq:relaxation_times_large} via Padé interpolation. (c) Curves show average position for a few quantum-jump trajectories of the unraveled Lindblad equation, starting from site 3.  Background shows heatmap of site occupancies, $n_j$, from the full Lindblad evolution.   (d) Perturbative estimates from Eq.'s (\ref{eq:relaxation_times_small}, \ref{eq:relaxation_times_large}) compared to numerical results obtained from Lindbladian spectrum. For panels (b,c,d) we used $L=7$ sites, $\varepsilon/\Delta = 1.0$, $J/\Delta=0.15$.  For (b,c), $\gamma/\Delta=  1.0$. 
}
\label{fig:energies_localized}
\end{figure}

{\em Model} ---  We demonstrate our mechanism using a bosonic tight-binding chain.  The Hamiltonian is
\begin{align}\label{eq:Hamiltonian}
    H = \sum_{j=2}^{L-1} J \left(b_{j+1}^\dagger b_j + \text{h.c.}\right) + \sum_{j=1}^L V_j b_j^\dagger b_j  .
\end{align}
Here $b_j$, $b_j^{\dagger}$ are bosonic operators for site $j$.  Site $1$ is isolated, with $V_1=0$.   
The remaining sites are subject to a linear (`Stark') potential, $V_{j\geq2}=\varepsilon + \Delta(L-j)$.  Here $\Delta$ is the slope of the potential and $\varepsilon$ is the potential gap between the first and last site (Fig.\ \ref{fig:energies_localized}(a)).  By using large enough $\varepsilon$, we ensure that the eigenstate $\ket{\psi_1}$ localized at the isolated site $1$ is in fact the ground state.   

The hopping  from site $2$ to $1$ does not appear in $H$ but is described by the jump operator $\Gamma = \sqrt{\gamma} b_{1}^\dagger b_{2}$ in the Lindbladian.  The full dissipative dynamics is described by a Lindblad master equation \cite{Lindblad_1976, GKS_1976, BreuerBook} $\dot{\rho} = \mathcal{L}(\rho)$ with the Lindbladian composed of the Hamiltonian (\ref{eq:Hamiltonian}) and jump operator $\Gamma$.  
The incoherent hopping `cools' the system, by driving particles to site 1 and hence to the ground state.  The Hamiltonian ground state localized on site 1 is also the unique steady state of the open system dynamics.

The eigenstates of the Hamiltonian (\ref{eq:Hamiltonian}) experience Wannier-Stark localization \cite{Wannier1962Localization}: we denote the eigenstate localized about site $j$ as $\ket{\psi_j}$.    If $\Delta/J$ is not too small, the localization length is small compared to the system size, and  the eigenenergies closely follow the potential landscape defined by  $V_j$.  This `energy landscape' has two local minima at the ends of the chain with an energy barrier in between given by the maximum at site 2.  We use this structure to realize an Mpemba effect. 

The mechanism is robust against various modifications, e.g., one could add a Hamiltonian hopping between sites 1 and 2, or several sites on the dissipative (left) part of the chain.  For concreteness, we describe the physics in terms of the minimal setup.  For simplicity, we assume the limit of strong localization; however in practice we find our analysis to be robust up to $J/\Delta \sim O(1)$.  

{\em Hierarchy of relaxation scales from Wannier-Stark localization} --- 
We first analyze single particle states that are initially exactly localized, $|\psi(0)\rangle = b_j^\dagger |0\rangle$. The resulting dynamics displays a striking (super-exponential!) separation of decay timescales corresponding to different starting positions, Fig.\ \ref{fig:energies_localized}(b).  This hierarchy of relaxation times is a result of the Stark localization: the Hamiltonian eigenstate localized at the starting site has a weight at site 2 that decreases (super-) exponentially with distance from site 2.  This weight determines the strength of coupling to the dissipation and hence the time scale, which is tunable via $J/\Delta$.  Since energy decreases with increasing $(j-2)$, high-energy initial states decay to the ground state faster, as seen in the pronounced crossing of curves in Fig.\ \ref{fig:energies_localized}(b)---a dramatic Mpemba effect.  

The relaxation times can be estimated in a quantum trajectory picture, obtained by unraveling the Lindblad master equation into an ensemble of stochastically evolving pure-state trajectories \cite{BreuerBook}. A few example trajectories are depicted in Fig. \ref{fig:energies_localized}(c). In the absence of quantum jumps the trajectories evolve deterministically under the non-Hermitian effective Hamiltonian $H_\text{nH} = H - \frac{i}{2}\Gamma^\dagger \Gamma = H - \frac{i}{2}\gamma n_2(1+n_1)$.  A particle initiated at a site $j>2$ mainly performs Bloch oscillations around its intial position; the eigenfunctions participating in this dynamics have small weight on site 2 and thus are only weakly affected by the non-Hermitian (loss) term in $H_\text{nH}$, seen in a very slow drift of the average position (away from site 2).  A quantum jump takes place when the particle is detected at site 2 and immediately taken to site 1, which is the steady state.  For small $\gamma$ and strong localization, the detection probability is given by the rate $\gamma$ times the weight of the Hamiltonian eigenstate $\ket{\psi_j}$ on site 2.   Hence the relaxation timescale is (End Matter)
\begin{align}\label{eq:relaxation_times_small}
    \tau_j \approx \frac{1}{\gamma} \frac{1}{\vert \psi_j(x=2)\vert^2} \approx \frac{1}{\gamma} [(j-2)!]^2 (\Delta/J)^{2(j-2)} 
\end{align}
for $\gamma \ll \Delta$. At stronger dissipation, the non-unitary dynamics of site 2 becomes important. For  $\gamma \gg \Delta$, we can estimate $\tau_j$ by tracing out site 2, yielding an effective master equation for the dynamics on sites $1, 3, 4, \dots, L$ with an effective incoherent hopping between sites 3 and 1 of strength $\gamma_\text{eff} = \frac{4J^2}{\gamma} \propto \frac{1}{\gamma}$ \cite{Zanardi2014_StrongDissPT, Popkov2018_StrongDissPT} (End Matter). Since $1/\gamma$ is a small parameter one can use the weak-dissipation result, replacing site 2 with site 3 and $\gamma$ with $\gamma_\text{eff}$, 
\begin{align}\label{eq:relaxation_times_large}
    \tau_j \approx \frac{\gamma}{4J^2} \frac{1}{\vert \psi_j(x=3)\vert^2} \approx \frac{\gamma}{4J^2} [(j-3)!]^2 (\Delta/J)^{2(j-3)} 
\end{align}
for $\gamma \gg \Delta$. 
The lifetimes can also be obtained from Lindbladian eigenvalues $\lambda_{jj}$ (discussed below).  Fig.\ \ref{fig:energies_localized}(d) compares the small-$\gamma$ and large-$\gamma$ expressions with numerical results from the Lindbladian spectrum.  The intermediate $\gamma$ regime is well-described by a Pad\'e interpolation between the limiting expressions above.

Our Mpemba mechanism is a purely quantum effect---as we demonstrate later, the corresponding classical setup has diffusive dynamics and does not display this parametric separation of timescales.

The analysis above is largely unaffected if we initiate the system in a localized energy eigenstate instead of a particular site.  The trajectories do not show Bloch oscillations, but otherwise the same Mpemba effect is seen.

{\em Liouvillian spectrum and eigenmodes} --- 
The time dependence of the density matrix, $\rho(t) = e^{\mathcal{L} t} (\rho(0))$, can be decomposed as $\rho(t) = \sum_\mu e^{\lambda_\mu t} \text{tr}[l_\mu^\dagger \rho(0)] r_\mu$, where $\lambda_\mu$ are the complex eigenvalues of the Lindbladian  and $l_\mu, r_\mu$ are its left and right eigenmatrices. The unique steady state $r_0 =: \rho_\infty$ with $\lambda_0 = 0$ will always be the state $\rho_\infty = |\psi_1\rangle\langle\psi_1|$, where $|\psi_1\rangle$ is the Hamiltonian eigenstate that is localized on the first site.

\begin{figure}
\resizebox{\columnwidth}{!}{\includegraphics{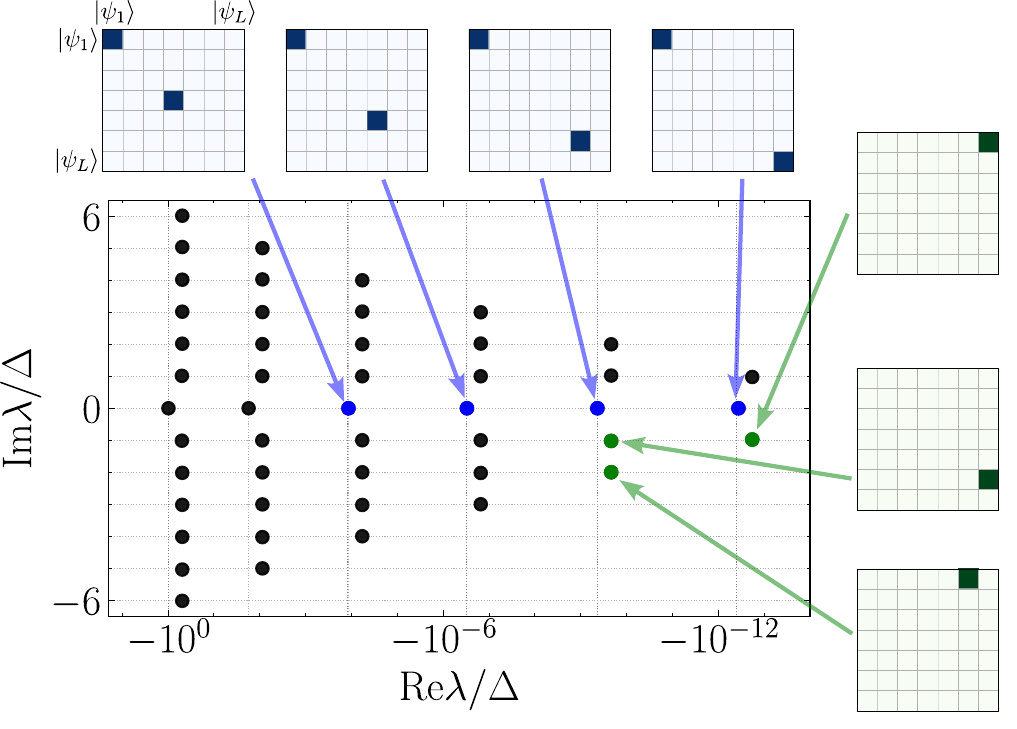}}
\caption{Lindbladian spectrum, with $J/\Delta=0.15, \varepsilon/\Delta = \gamma/\Delta = 1.0$.  Vertical grid lines show $-1/\tau_j$, with $\tau_j$ estimated by Padé interpolation of Eq. (\ref{eq:relaxation_times_small}, \ref{eq:relaxation_times_large}).  Square panels show absolute values of the matrix elements of selected right eigenmodes in the energy eigenbasis.  }
\label{fig:spectrum}
\end{figure} 

Fig.\ \ref{fig:spectrum} shows the Lindbladian spectrum obtained numerically, as well as the matrix elements of a few selected right eigenmodes in the energy eigenbasis.  A remarkable feature of our setup is that the eigenmodes (except the fastest decaying ones) are extremely well localized in the energy eigenbasis.  This is due to Stark localization, which prevents the spatially restricted dissipation from strongly coupling different energy eigenstates (End Matter).  This unusual correspondence between Hamiltonian eigenstates and Liouvillian eigenmodes enables the parametric Mpemba effect in our system.  

Eigenmodes corresponding to complex eigenvalues are closely related to coherences between energy eigenstates, i.e., they have the form $r_{jk} \approx|\psi_j\rangle\langle\psi_k|, j \neq k$, whereas those corresponding to real eigenvalues are related to the populations, with $r_{jj} \approx |\psi_j\rangle\langle\psi_j| - |\psi_1\rangle\langle\psi_1|, \ j \geq 2$  (Fig.\ \ref{fig:spectrum} and End Matter). As a result, the corresponding eigenvalues $\lambda_{jj}$ are directly connected to decay timescales $\tau_j$ of states initiated in eigenstates $\ket{\psi_j}$ (or on sites $j$), namely, $\tau_j=1/|\lambda_{jj}|$.  This allows us to extract timescales numerically from the Liouvlillian spectrum,   Fig.\ \ref{fig:energies_localized}(d).  The hierarchy of relaxation timescales $\tau_j$ corresponds to a hierarchy of metastable eigenmodes \cite{Macieszczak2016_OpenMetastability, Macieszczak2021_ClassicalMetastability}.

The spectrum is highly structured: the eigenvalues form vertical stripes, with the real parts very close to $- \frac{1}{\tau_j}$ and $- \smash{\frac{1}{2} \frac{1}{\tau_j}}$ for real and complex eigenvalues respectively. The complex eigenvalues represent coherences between $\ket{\psi_j}$ and other energy eigenstates that decay slower, i.e., $\ket{\psi_j}\bra{\psi_k}$ with $k>j$ and $k=1$. These have the decay rates $\frac{1}{2}|\lambda_{jj} + \lambda_{kk}| \approx |\lambda_{jj}|/2$. 
The imaginary part of the complex eigenvalues is roughly given by the energy difference of the states that make up the coherence, $E_j-E_k$, which in the localized regime is close to the potential spacing $\Delta$ or the gap $\varepsilon$ between the first and last sites.

We consider initial states (sites, energy eigenstates, Gibbs states) that are (close to) diagonal in the  Hamiltonian eigenbasis: as a result the dynamics is determined by the eigenmodes corresponding to real eigenvalues.  Of particular relevance for the Mpemba effect is the eigenmode decaying at the longest timescale; the corresponding left eigenmode is    $l_\text{SM} \approx |\psi_L\rangle\langle \psi_L|$.

\begin{figure}
\resizebox{\columnwidth}{!}{\includegraphics{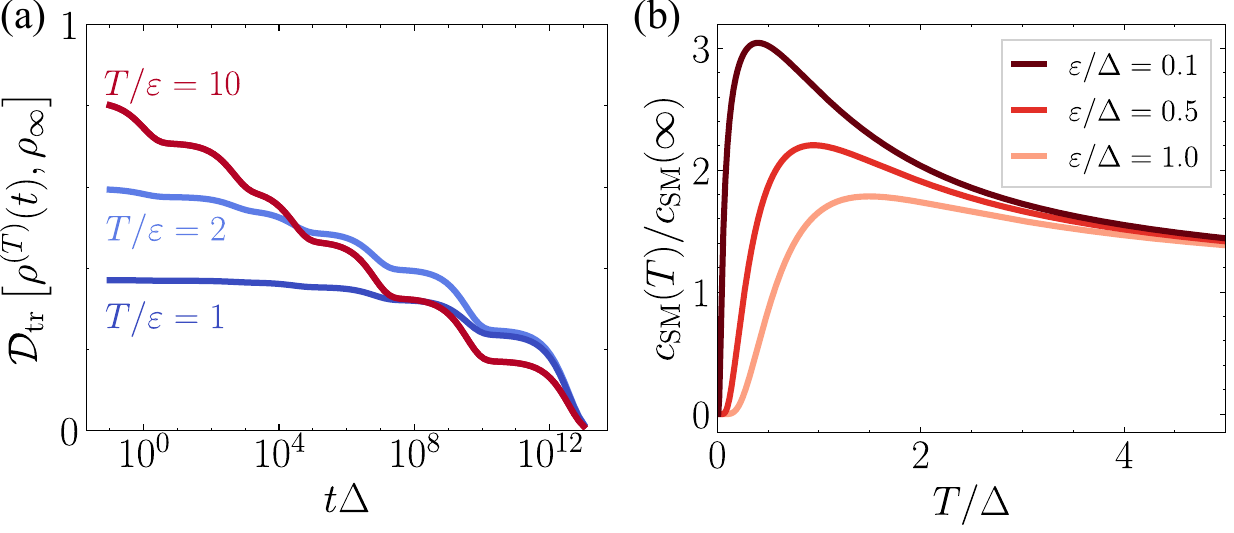}}
\caption{\label{fig:thermal_Mpemba}
Initiating with Gibbs states $\rho^{(T)}$. 
(a) Trace distance to the steady state $\rho_\infty = |\psi_1\rangle\langle\psi_1|$ as function of time.  The curve crossings represent the Mpemba effect.  (b) Scaled overlap  of Gibbs states with the relevant slowest decaying left eigenmode.  Non-monotonic $T$-dependence is a signature of the Mpemba effect.  In both panels: $L=7$, $J/\Delta = 0.15$, $\gamma/\Delta = 1.0$.  For panel (a), $\varepsilon/\Delta = 1.0$. 
}
\end{figure}

{\em Thermal initial states} --- 
We now consider Gibbs states $\rho^{(T)} \propto e^{-H/T}$ as initial states, in order to connect more closely to the original context of the Mpemba effect.  We find a wide temperature regime for which hotter initial states cool down faster than colder initial states. 
This is because Gibbs states with $T \gtrsim \varepsilon$ have less and less weight in the slow-decaying low-energy states on the right end of the chain the more $T$ is increased.  
In Fig.\ \ref{fig:thermal_Mpemba}(a) we plot the trace distance 
$\frac{1}{2} \text{tr}|\rho^{(T)}(t) - \rho_\infty|$ \cite{NielsenChuang_2010, BreuerBook} of the time evolved state $\rho^{(T)}(t) := e^{\mathcal{L}t} (\rho^{(T)})$ to the steady state $\rho_\infty$. 

This distance measure decreases monotonically under Lindbladian time evolution \cite{BreuerBook}, making it an unambiguous measure for identifying Mpemba effects \cite{Lu_Raz_pnas2017_Mpemba}.  The curves in Fig.\  \ref{fig:thermal_Mpemba}(a) show pronounced Mpemba crossings.

Another indicator of the Mpemba effect is the overlap of the slowest decaying mode with thermal initial states, $c_\text{SM}(T) = \text{tr}\big[l_\text{SM}^\dagger \rho^{(T)}\big]$.  A non-monotonic variation of $c_\text{SM}$ with $T$ is a signature of an Mpemba effect \cite{Lu_Raz_pnas2017_Mpemba}.  Fig.\ \ref{fig:thermal_Mpemba}(b) shows that, in our case, the overlap increases with $T$ until a maximum at $T \approx \mathcal{O}(\varepsilon)$, corresponding roughly to the energy of the slowest decaying low-energy mode $\ket{\psi_L}$,  localized on the right end of the chain.  Further increase in temperature leads to a decreasing occupation of this mode, so the overlap decreases again, approaching $c_\text{SM}\to1/L$ for $T\to\infty$.  The peak height increases with decreasing $\varepsilon$, reaching $c_\text{SM}=1/2$ for small $\varepsilon$ when $\ket{\psi_1}$ and $\ket{\psi_L}$ are almost degenerate.   
Thus, both the peak location and the peak height can be tuned via $\varepsilon$, affording control over the prominence of the Mpemba effect.

\begin{figure}
\resizebox{\columnwidth}{!}{\includegraphics{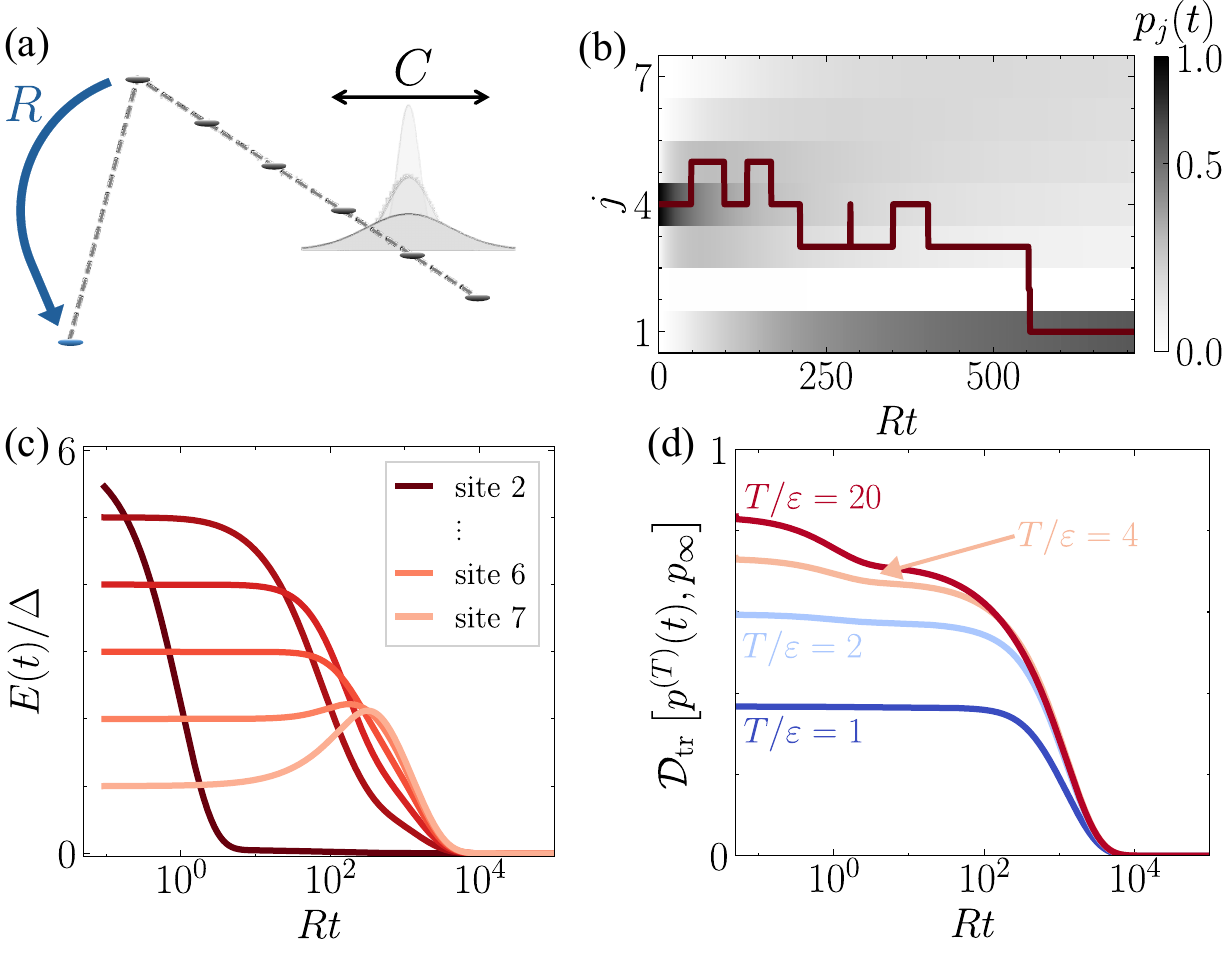}}
\caption{Corresponding classical Markov process. (a) Setup, with transition rates $R, C$.  Simulation results are for  $R/C = 100$. (b) Particle position for a single Gillespie trajectory; background shows the position distribution $p_j(t)$ obtained from the rate equation for the same initial condition.  Diffusive spreading is visible for $j>2$.  (c) Time evolution of energy $E(t) := \sum_j p_j(t)V_j$, showing different initial states have comparable lifetimes.  (d) Trace distance of a time evolved Gibbs state to the steady state for different temperatures. Mpemba crossings are very mild if present at all.
}
\label{fig:classical}
\end{figure}

{\em Classical analog} --- To illustrate the quantum nature of the Mpemba effect, we compare the Lindbladian dynamics to those of a suitable classical Markovian process \cite{vanKampen1992_book}. The particle occupation is described by a probability distribution $p_j(t)$, $1 \leq j \leq L$, which evolves under a master equation $\dot{\vec{p}}(t) = \mathcal{K} \vec{p}(t)$, where $\mathcal{K}$ is the Markov (Kolmogorov) generator of the dynamics (End Matter). The master equation allows for transitions between neighboring sites with a rate $C$ for $j \geq 2$, and a transition from site 2 to site 1 with a rate $R \gg C$.  The steady state is then exactly localized on site 1.  To each site we associate an energy given by the potential landscape $\smash{\{V_j\}}$, so that the total energy is $E(t) = \sum_{j=1}^L p_j(t)V_j$.

As shown in Fig.\ \ref{fig:classical}(b),  initially localized particle distributions spread out diffusively, in stark contrast to the localized Bloch oscillations in the quantum case [Fig.~\ref{fig:energies_localized}(c)]. As the spreading is diffusive, we expect the lifetime to scale as $\tau_{j} \propto (j-2)^2$, which gives $\tau_{j+1}/\tau_{j} \to 1$ for large $j$. Thus, the Mpemba effect should vanish except for small system sizes.  Fig.\  \ref{fig:classical}(c) indeed shows that states initially localized away from site 2 decay on comparable time scales. 
Fig. \ref{fig:classical}(d) shows the relaxation of thermal initial states, lacking Mpemba-like crossings that are very pronounced in the quantum case [Fig.~\ref{fig:thermal_Mpemba}(a)].

\begin{figure}
\resizebox{\columnwidth}{!}{\includegraphics{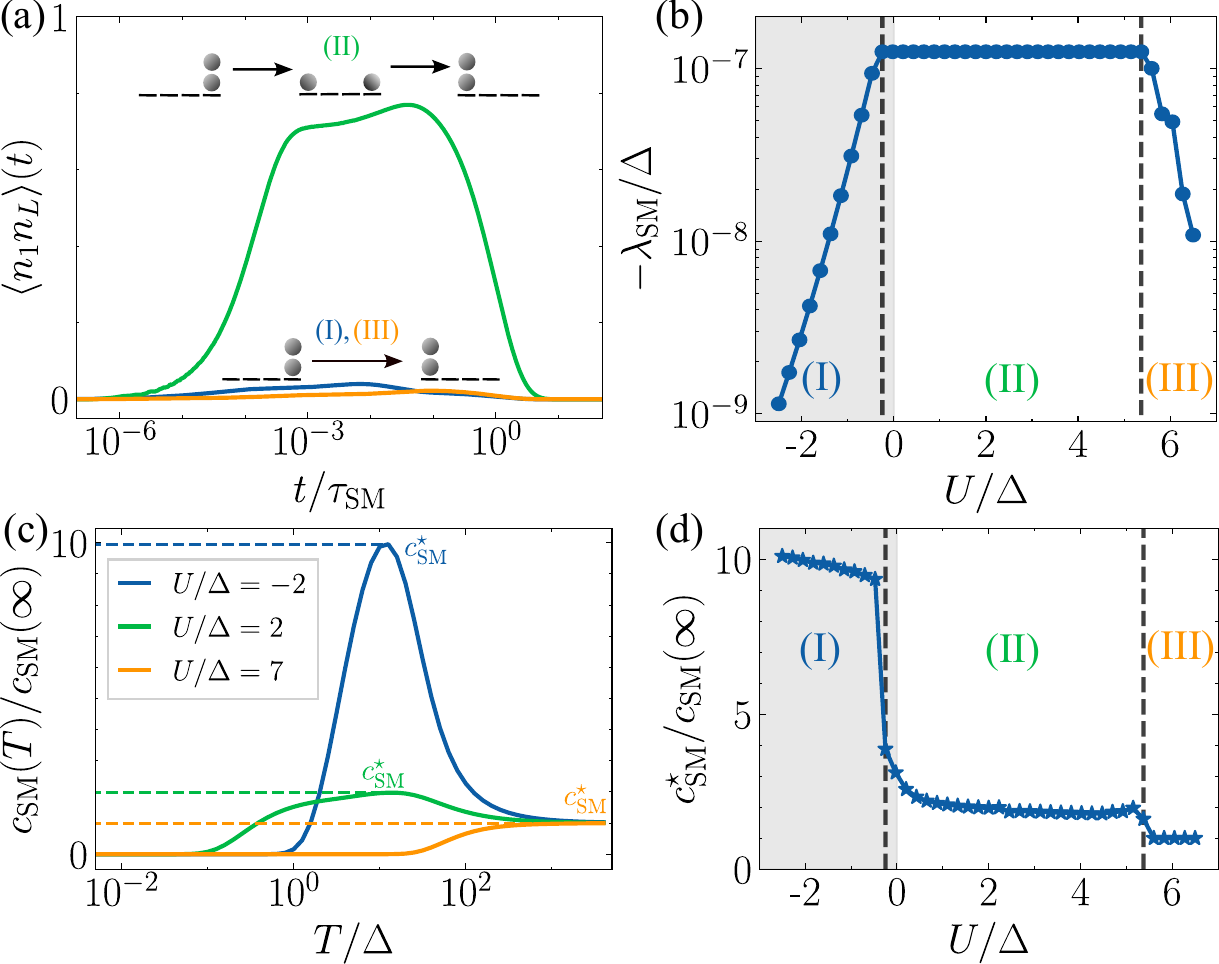}}
\caption{Two bosons with varying on-site interaction $U$ with $L=7$, $J/\Delta=0.5$, $\varepsilon/\Delta=0.25$, $\gamma/\Delta = 10$. 
(a) Time evolution from the initial state $\frac{1}{\sqrt{2}}(b_L^\dagger)^2|0\rangle$, where $\tau_\text{SM} := 1 / \vert \lambda_\text{SM}(U)\vert$ is the decay time of the slowest mode with nonzero overlap.  The large values of $\xpct{n_1 n_L}$ for intermediate $U$ (regime II) indicate a two-step process where the particles reach site 1 sequentially.  In contrast, the relaxation is collective in regimes (I) and (III). 
(b) The gap is unaffected by $U$ in (II) and decreases with $|U|$ in (I), (III). %
(c) Overlap of Gibbs states with the slow mode.  The non-monotonicity (Mpemba effect) is greatly enhanced by attractive interactions in (I), and suppressed by strong repulsive interactions in (III).  
This is reflected in the variation of the peak overlap shown in (d). We use $U_1 = U$ for $U < 0$ (shaded) and $U_1 = 0$ for $U > 0$ (white).
}
\label{fig:two_particles}
\end{figure}

{\em Interacting bosons} --- 
Next, we consider two bosons subject to the same potential landscape \eqref{eq:Hamiltonian} and jump operator $\Gamma = \smash{\sqrt{\gamma} b_{1}^\dagger b_{2}}$ .  In addition, we have an on-site (Bose-Hubbard) interaction $H_\text{BH} = \frac{1}{2} \sum_{j=1}^L U_j n_j (n_j-1)$.
We choose a uniform interaction strength $U_j = U$ on the downward slope of the potential $j \geq 2$, and adjust the interaction strength on the first site such that the ground state of the Hamiltonian remains localized on that site, setting $U_1=0$ for $U>0$ and $U_1 = U$ for $U<0$.

Few-particle interacting systems on Stark ladders can exhibit localization and Bloch oscillations \cite{DiasEtAl_PRB2007_2electronBlochOscil, Khomeriki_Krimer_Haque_Flach_PRA2010, Longhi_DellaValle_PRB2012_2anyonBlochOscil, Meinert_Naegerl_PRL2014_BlochOscilExpt, Preiss_Greiner_Science2015_quantumwalks, Wiater_Zakrzewski_PRA2017_2bosons, RibeiroLazaridesHaque_PRL2020, Sarkar_Sowinski_PRA2020, Zhang_EtAl_PRB2024_2doublonBlochOscil}; the eigenstates can have particles localized on the same or on different sites.  This allows richer versions of our Mpemba mechanism to persist in the interacting system.  

Interaction effects are summarized in Fig.\ \ref{fig:two_particles}.  There are three regimes: In regime (I) ($U$ negative up to a value close to zero) and in regime (III) (large and positive $U$) the relaxation occurs through a doublon pathway; the pair of particles decays collectively, whereas  in the intermediate regime (II), they decay separately.  This difference is strikingly manifested in the time evolution of the correlator $\langle n_1 n_L\rangle(t)$, Fig. \ref{fig:two_particles}(a). 

The slowest decaying real eigenmode of the Lindbladian changes character at the two regime boundaries  (I)--(II) and (II)--(III).  Fig.\ \ref{fig:two_particles}(b) shows sharp changes in the real gap.  In the regimes of collective decay, (I) and (III), the slowest real eigenmode corresponds to the Hamiltonian eigenstate having both particles localized as a doublon on  site $L$.  Since the Hubbard interaction suppresses transitions out of this state, the lifetime of this mode  increases with increasing interaction strength $|U|$.  The gap therefore decreases for both $U\to\infty$ and $U\to-\infty$, Fig.\ \ref{fig:two_particles}(b).  In the intermediate regime II, the slowest real mode instead corresponds to one particle localized at site $1$ and the other particle at site $L$. 
The decay of this mode is not affected by the interaction, hence the gap is constant.

Figs.\ \ref{fig:two_particles}(c,d) show that the Mpemba effect, measured by the height of the peak of the overlap curve $c_\text{SM}(T)$, is enormously enhanced  in regime (I), is moderate in regime (II), and absent in regime (III).  

In regime (I), the slowest mode is energetically close to the ground state (gap $\sim\varepsilon$), so that the overlap with a thermal state is maximized for a temperature comparable to this
gap. Further increase in temperature reduces the overlap all the way to $1/D$. So one gets a sharp peak.    In regime (III), on the other hand, the same slow mode has a high energy, such that the overlap with a thermal state continuously grows with $T$, approaching $1/D$ for $T\to\infty$.  There is thus no peak and no Mpemba.

{\em Summary and outlook} ---
We formulated a mechanism to engineer a parametric quantum Mpemba effect by combining tilted potentials that provide an energy landscape and cooling via local incoherent hops. 
We further demonstrated how the interplay of localization and coherent interactions can create additional metastable states whose energy can be tuned to shape the Mpemba effect. 

Our mechanism does not require fine tuning of model parameters, so long as the energy eigenstates are sufficiently well localized.
Increasing the localization length can drastically decrease the timescales while keeping the mechanism intact. For example, with $J/\Delta = 0.5$, all the crossings in Figs.~\ref{fig:energies_localized}(b) and \ref{fig:thermal_Mpemba}(a) occur within 
$\sim 10^5$ Bloch oscillations, which might be in reach of current cold-atom experiments \cite{Ferrari2006, Gustavsson2008}.
The incoherent hops may be realized by laser-assisted tunneling to an excited state followed by spontaneous emission  \cite{Haga_Ueda_PRL2021_LiouvillianSkinEffect, Sharma2021, Garbe2024}, or by local measurement plus conditional feedback \cite{wiseman2009book}. While the protocol in Ref.~\cite{Haga_Ueda_PRL2021_LiouvillianSkinEffect} also gives rise to dephasing at these sites, we have checked that our results are virtually unaffected under such dephasing.

\begin{acknowledgments}
NA and MH acknowledge support from the Deutsche Forschungsgemeinschaft under grant SFB 1143 (project-id 247310070).  
\end{acknowledgments}

\bibliography{refs}

\begin{thebibliography}{69}%
\makeatletter
\providecommand \@ifxundefined [1]{%
 \@ifx{#1\undefined}
}%
\providecommand \@ifnum [1]{%
 \ifnum #1\expandafter \@firstoftwo
 \else \expandafter \@secondoftwo
 \fi
}%
\providecommand \@ifx [1]{%
 \ifx #1\expandafter \@firstoftwo
 \else \expandafter \@secondoftwo
 \fi
}%
\providecommand \natexlab [1]{#1}%
\providecommand \enquote  [1]{``#1''}%
\providecommand \bibnamefont  [1]{#1}%
\providecommand \bibfnamefont [1]{#1}%
\providecommand \citenamefont [1]{#1}%
\providecommand \href@noop [0]{\@secondoftwo}%
\providecommand \href [0]{\begingroup \@sanitize@url \@href}%
\providecommand \@href[1]{\@@startlink{#1}\@@href}%
\providecommand \@@href[1]{\endgroup#1\@@endlink}%
\providecommand \@sanitize@url [0]{\catcode `\\12\catcode `\$12\catcode `\&12\catcode `\#12\catcode `\^12\catcode `\_12\catcode `\%12\relax}%
\providecommand \@@startlink[1]{}%
\providecommand \@@endlink[0]{}%
\providecommand \url  [0]{\begingroup\@sanitize@url \@url }%
\providecommand \@url [1]{\endgroup\@href {#1}{\urlprefix }}%
\providecommand \urlprefix  [0]{URL }%
\providecommand \Eprint [0]{\href }%
\providecommand \doibase [0]{https://doi.org/}%
\providecommand \selectlanguage [0]{\@gobble}%
\providecommand \bibinfo  [0]{\@secondoftwo}%
\providecommand \bibfield  [0]{\@secondoftwo}%
\providecommand \translation [1]{[#1]}%
\providecommand \BibitemOpen [0]{}%
\providecommand \bibitemStop [0]{}%
\providecommand \bibitemNoStop [0]{.\EOS\space}%
\providecommand \EOS [0]{\spacefactor3000\relax}%
\providecommand \BibitemShut  [1]{\csname bibitem#1\endcsname}%
\let\auto@bib@innerbib\@empty
\bibitem [{\citenamefont {Macieszczak}\ \emph {et~al.}(2016)\citenamefont {Macieszczak}, \citenamefont {Gu{\c{t}}{\u{a}}}, \citenamefont {Lesanovsky},\ and\ \citenamefont {Garrahan}}]{Macieszczak2016_OpenMetastability}%
  \BibitemOpen
  \bibfield  {author} {\bibinfo {author} {\bibfnamefont {K.}~\bibnamefont {Macieszczak}}, \bibinfo {author} {\bibfnamefont {M.}~\bibnamefont {Gu{\c{t}}{\u{a}}}}, \bibinfo {author} {\bibfnamefont {I.}~\bibnamefont {Lesanovsky}},\ and\ \bibinfo {author} {\bibfnamefont {J.~P.}\ \bibnamefont {Garrahan}},\ }\bibfield  {title} {\bibinfo {title} {Towards a theory of metastability in open quantum dynamics},\ }\href {https://doi.org/10.1103/PhysRevLett.116.240404} {\bibfield  {journal} {\bibinfo  {journal} {Phys. Rev. Lett.}\ }\textbf {\bibinfo {volume} {116}},\ \bibinfo {pages} {240404} (\bibinfo {year} {2016})}\BibitemShut {NoStop}%
\bibitem [{\citenamefont {Minganti}\ \emph {et~al.}(2018)\citenamefont {Minganti}, \citenamefont {Biella}, \citenamefont {Bartolo},\ and\ \citenamefont {Ciuti}}]{Minganti2018}%
  \BibitemOpen
  \bibfield  {author} {\bibinfo {author} {\bibfnamefont {F.}~\bibnamefont {Minganti}}, \bibinfo {author} {\bibfnamefont {A.}~\bibnamefont {Biella}}, \bibinfo {author} {\bibfnamefont {N.}~\bibnamefont {Bartolo}},\ and\ \bibinfo {author} {\bibfnamefont {C.}~\bibnamefont {Ciuti}},\ }\bibfield  {title} {\bibinfo {title} {Spectral theory of {L}iouvillians for dissipative phase transitions},\ }\href {https://doi.org/10.1103/physreva.98.042118} {\bibfield  {journal} {\bibinfo  {journal} {Phys. Rev. A}\ }\textbf {\bibinfo {volume} {98}},\ \bibinfo {pages} {042118} (\bibinfo {year} {2018})}\BibitemShut {NoStop}%
\bibitem [{\citenamefont {Macieszczak}\ \emph {et~al.}(2021)\citenamefont {Macieszczak}, \citenamefont {Rose}, \citenamefont {Lesanovsky},\ and\ \citenamefont {Garrahan}}]{Macieszczak2021_ClassicalMetastability}%
  \BibitemOpen
  \bibfield  {author} {\bibinfo {author} {\bibfnamefont {K.}~\bibnamefont {Macieszczak}}, \bibinfo {author} {\bibfnamefont {D.~C.}\ \bibnamefont {Rose}}, \bibinfo {author} {\bibfnamefont {I.}~\bibnamefont {Lesanovsky}},\ and\ \bibinfo {author} {\bibfnamefont {J.~P.}\ \bibnamefont {Garrahan}},\ }\bibfield  {title} {\bibinfo {title} {Theory of classical metastability in open quantum systems},\ }\href {https://doi.org/10.1103/PhysRevResearch.3.033047} {\bibfield  {journal} {\bibinfo  {journal} {Phys. Rev. Res.}\ }\textbf {\bibinfo {volume} {3}},\ \bibinfo {pages} {033047} (\bibinfo {year} {2021})}\BibitemShut {NoStop}%
\bibitem [{\citenamefont {Rose}\ \emph {et~al.}(2022)\citenamefont {Rose}, \citenamefont {Macieszczak}, \citenamefont {Lesanovsky},\ and\ \citenamefont {Garrahan}}]{Rose2022}%
  \BibitemOpen
  \bibfield  {author} {\bibinfo {author} {\bibfnamefont {D.~C.}\ \bibnamefont {Rose}}, \bibinfo {author} {\bibfnamefont {K.}~\bibnamefont {Macieszczak}}, \bibinfo {author} {\bibfnamefont {I.}~\bibnamefont {Lesanovsky}},\ and\ \bibinfo {author} {\bibfnamefont {J.~P.}\ \bibnamefont {Garrahan}},\ }\bibfield  {title} {\bibinfo {title} {Hierarchical classical metastability in an open quantum {E}ast model},\ }\href {https://doi.org/10.1103/physreve.105.044121} {\bibfield  {journal} {\bibinfo  {journal} {Phys. Rev. E}\ }\textbf {\bibinfo {volume} {105}},\ \bibinfo {pages} {044121} (\bibinfo {year} {2022})}\BibitemShut {NoStop}%
\bibitem [{\citenamefont {Zhang}\ \emph {et~al.}(2025{\natexlab{a}})\citenamefont {Zhang}, \citenamefont {Jin}, \citenamefont {Qiu}, \citenamefont {Ma},\ and\ \citenamefont {Liu}}]{Zhang2025_metastability}%
  \BibitemOpen
  \bibfield  {author} {\bibinfo {author} {\bibfnamefont {J.-X.}\ \bibnamefont {Zhang}}, \bibinfo {author} {\bibfnamefont {Y.-D.}\ \bibnamefont {Jin}}, \bibinfo {author} {\bibfnamefont {C.-D.}\ \bibnamefont {Qiu}}, \bibinfo {author} {\bibfnamefont {W.-L.}\ \bibnamefont {Ma}},\ and\ \bibinfo {author} {\bibfnamefont {G.-Q.}\ \bibnamefont {Liu}},\ }\bibfield  {title} {\bibinfo {title} {Observation of metastability in open quantum dynamics of a solid-state system},\ }\href {https://doi.org/10.1038/s41467-025-64772-6} {\bibfield  {journal} {\bibinfo  {journal} {Nat. Commun.}\ }\textbf {\bibinfo {volume} {16}},\ \bibinfo {pages} {9818} (\bibinfo {year} {2025}{\natexlab{a}})}\BibitemShut {NoStop}%
\bibitem [{\citenamefont {Bechhoefer}\ \emph {et~al.}(2021)\citenamefont {Bechhoefer}, \citenamefont {Kumar},\ and\ \citenamefont {Ch{\'e}trite}}]{Bechhoefer2021}%
  \BibitemOpen
  \bibfield  {author} {\bibinfo {author} {\bibfnamefont {J.}~\bibnamefont {Bechhoefer}}, \bibinfo {author} {\bibfnamefont {A.}~\bibnamefont {Kumar}},\ and\ \bibinfo {author} {\bibfnamefont {R.}~\bibnamefont {Ch{\'e}trite}},\ }\bibfield  {title} {\bibinfo {title} {A fresh understanding of the {M}pemba effect},\ }\href {https://doi.org/10.1038/s42254-021-00349-8} {\bibfield  {journal} {\bibinfo  {journal} {Nat. Rev. Phys.}\ }\textbf {\bibinfo {volume} {3}},\ \bibinfo {pages} {534} (\bibinfo {year} {2021})}\BibitemShut {NoStop}%
\bibitem [{\citenamefont {Teza}\ \emph {et~al.}(2026)\citenamefont {Teza}, \citenamefont {Bechhoefer}, \citenamefont {Lasanta}, \citenamefont {Raz},\ and\ \citenamefont {Vucelja}}]{Teza2026}%
  \BibitemOpen
  \bibfield  {author} {\bibinfo {author} {\bibfnamefont {G.}~\bibnamefont {Teza}}, \bibinfo {author} {\bibfnamefont {J.}~\bibnamefont {Bechhoefer}}, \bibinfo {author} {\bibfnamefont {A.}~\bibnamefont {Lasanta}}, \bibinfo {author} {\bibfnamefont {O.}~\bibnamefont {Raz}},\ and\ \bibinfo {author} {\bibfnamefont {M.}~\bibnamefont {Vucelja}},\ }\bibfield  {title} {\bibinfo {title} {Speedups in nonequilibrium thermal relaxation: {M}pemba and related effects},\ }\href {https://doi.org/10.1016/j.physrep.2025.10.009} {\bibfield  {journal} {\bibinfo  {journal} {Phys. Rep.}\ }\textbf {\bibinfo {volume} {1164}},\ \bibinfo {pages} {1} (\bibinfo {year} {2026})}\BibitemShut {NoStop}%
\bibitem [{\citenamefont {Ares}\ \emph {et~al.}(2025)\citenamefont {Ares}, \citenamefont {Calabrese},\ and\ \citenamefont {Murciano}}]{Ares_Calabrese_Murciano_NatRevPhys2025_Mpemba_review}%
  \BibitemOpen
  \bibfield  {author} {\bibinfo {author} {\bibfnamefont {F.}~\bibnamefont {Ares}}, \bibinfo {author} {\bibfnamefont {P.}~\bibnamefont {Calabrese}},\ and\ \bibinfo {author} {\bibfnamefont {S.}~\bibnamefont {Murciano}},\ }\bibfield  {title} {\bibinfo {title} {{The quantum Mpemba effects}},\ }\href {https://doi.org/10.1038/s42254-025-00838-0} {\bibfield  {journal} {\bibinfo  {journal} {Nat. Rev. Phys.}\ }\textbf {\bibinfo {volume} {7}},\ \bibinfo {pages} {451} (\bibinfo {year} {2025})}\BibitemShut {NoStop}%
\bibitem [{\citenamefont {Calabrese}(2026)}]{Calabrese2026}%
  \BibitemOpen
  \bibfield  {author} {\bibinfo {author} {\bibfnamefont {P.}~\bibnamefont {Calabrese}},\ }\bibfield  {title} {\bibinfo {title} {The quantum {M}pemba effect in closed systems: from theory to experiment},\ }\href {https://doi.org/10.1088/1742-5468/ae4bb6} {\bibfield  {journal} {\bibinfo  {journal} {J. Stat. Mech.}\ }\textbf {\bibinfo {volume} {2026}},\ \bibinfo {pages} {034002} (\bibinfo {year} {2026})}\BibitemShut {NoStop}%
\bibitem [{\citenamefont {Ares}\ \emph {et~al.}(2023)\citenamefont {Ares}, \citenamefont {Murciano},\ and\ \citenamefont {Calabrese}}]{Ares2023}%
  \BibitemOpen
  \bibfield  {author} {\bibinfo {author} {\bibfnamefont {F.}~\bibnamefont {Ares}}, \bibinfo {author} {\bibfnamefont {S.}~\bibnamefont {Murciano}},\ and\ \bibinfo {author} {\bibfnamefont {P.}~\bibnamefont {Calabrese}},\ }\bibfield  {title} {\bibinfo {title} {Entanglement asymmetry as a probe of symmetry breaking},\ }\href {https://doi.org/10.1038/s41467-023-37747-8} {\bibfield  {journal} {\bibinfo  {journal} {Nat. Commun.}\ }\textbf {\bibinfo {volume} {14}},\ \bibinfo {pages} {2036} (\bibinfo {year} {2023})}\BibitemShut {NoStop}%
\bibitem [{\citenamefont {Murciano}\ \emph {et~al.}(2024)\citenamefont {Murciano}, \citenamefont {Ares}, \citenamefont {Klich},\ and\ \citenamefont {Calabrese}}]{Murciano2024}%
  \BibitemOpen
  \bibfield  {author} {\bibinfo {author} {\bibfnamefont {S.}~\bibnamefont {Murciano}}, \bibinfo {author} {\bibfnamefont {F.}~\bibnamefont {Ares}}, \bibinfo {author} {\bibfnamefont {I.}~\bibnamefont {Klich}},\ and\ \bibinfo {author} {\bibfnamefont {P.}~\bibnamefont {Calabrese}},\ }\bibfield  {title} {\bibinfo {title} {Entanglement asymmetry and quantum {M}pemba effect in the {XY} spin chain},\ }\href {https://doi.org/10.1088/1742-5468/ad17b4} {\bibfield  {journal} {\bibinfo  {journal} {J. Stat. Mech.}\ }\textbf {\bibinfo {volume} {2024}},\ \bibinfo {pages} {013103} (\bibinfo {year} {2024})}\BibitemShut {NoStop}%
\bibitem [{\citenamefont {Rylands}\ \emph {et~al.}(2024)\citenamefont {Rylands}, \citenamefont {Klobas}, \citenamefont {Ares}, \citenamefont {Calabrese}, \citenamefont {Murciano},\ and\ \citenamefont {Bertini}}]{Rylands2024}%
  \BibitemOpen
  \bibfield  {author} {\bibinfo {author} {\bibfnamefont {C.}~\bibnamefont {Rylands}}, \bibinfo {author} {\bibfnamefont {K.}~\bibnamefont {Klobas}}, \bibinfo {author} {\bibfnamefont {F.}~\bibnamefont {Ares}}, \bibinfo {author} {\bibfnamefont {P.}~\bibnamefont {Calabrese}}, \bibinfo {author} {\bibfnamefont {S.}~\bibnamefont {Murciano}},\ and\ \bibinfo {author} {\bibfnamefont {B.}~\bibnamefont {Bertini}},\ }\bibfield  {title} {\bibinfo {title} {Microscopic origin of the quantum {M}pemba effect in integrable systems},\ }\href {https://doi.org/10.1103/physrevlett.133.010401} {\bibfield  {journal} {\bibinfo  {journal} {Phys. Rev. Lett.}\ }\textbf {\bibinfo {volume} {133}},\ \bibinfo {pages} {010401} (\bibinfo {year} {2024})}\BibitemShut {NoStop}%
\bibitem [{\citenamefont {Liu}\ \emph {et~al.}(2024)\citenamefont {Liu}, \citenamefont {Zhang}, \citenamefont {Yin},\ and\ \citenamefont {Zhang}}]{Liu2024}%
  \BibitemOpen
  \bibfield  {author} {\bibinfo {author} {\bibfnamefont {S.}~\bibnamefont {Liu}}, \bibinfo {author} {\bibfnamefont {H.-K.}\ \bibnamefont {Zhang}}, \bibinfo {author} {\bibfnamefont {S.}~\bibnamefont {Yin}},\ and\ \bibinfo {author} {\bibfnamefont {S.-X.}\ \bibnamefont {Zhang}},\ }\bibfield  {title} {\bibinfo {title} {Symmetry restoration and quantum {M}pemba effect in symmetric random circuits},\ }\href {https://doi.org/10.1103/physrevlett.133.140405} {\bibfield  {journal} {\bibinfo  {journal} {Phys. Rev. Lett.}\ }\textbf {\bibinfo {volume} {133}},\ \bibinfo {pages} {140405} (\bibinfo {year} {2024})}\BibitemShut {NoStop}%
\bibitem [{\citenamefont {Turkeshi}\ \emph {et~al.}(2025)\citenamefont {Turkeshi}, \citenamefont {Calabrese},\ and\ \citenamefont {De~Luca}}]{Turkeshi2025}%
  \BibitemOpen
  \bibfield  {author} {\bibinfo {author} {\bibfnamefont {X.}~\bibnamefont {Turkeshi}}, \bibinfo {author} {\bibfnamefont {P.}~\bibnamefont {Calabrese}},\ and\ \bibinfo {author} {\bibfnamefont {A.}~\bibnamefont {De~Luca}},\ }\bibfield  {title} {\bibinfo {title} {Quantum {M}pemba effect in random circuits},\ }\href {https://doi.org/10.1103/5d6p-8d1b} {\bibfield  {journal} {\bibinfo  {journal} {Phys. Rev. Lett.}\ }\textbf {\bibinfo {volume} {135}},\ \bibinfo {pages} {040403} (\bibinfo {year} {2025})}\BibitemShut {NoStop}%
\bibitem [{\citenamefont {Bhore}\ \emph {et~al.}(2025)\citenamefont {Bhore}, \citenamefont {Su}, \citenamefont {Martin}, \citenamefont {Clerk},\ and\ \citenamefont {Papi{\'c}}}]{Bhore2025}%
  \BibitemOpen
  \bibfield  {author} {\bibinfo {author} {\bibfnamefont {T.}~\bibnamefont {Bhore}}, \bibinfo {author} {\bibfnamefont {L.}~\bibnamefont {Su}}, \bibinfo {author} {\bibfnamefont {I.}~\bibnamefont {Martin}}, \bibinfo {author} {\bibfnamefont {A.~A.}\ \bibnamefont {Clerk}},\ and\ \bibinfo {author} {\bibfnamefont {Z.}~\bibnamefont {Papi{\'c}}},\ }\bibfield  {title} {\bibinfo {title} {Quantum {M}pemba effect without global symmetries},\ }\href {https://doi.org/10.1103/1td3-2vwf} {\bibfield  {journal} {\bibinfo  {journal} {Phys. Rev. B}\ }\textbf {\bibinfo {volume} {112}},\ \bibinfo {pages} {L121109} (\bibinfo {year} {2025})}\BibitemShut {NoStop}%
\bibitem [{\citenamefont {M\"{u}ller}\ \emph {et~al.}(2026)\citenamefont {M\"{u}ller}, \citenamefont {Pappalardi},\ and\ \citenamefont {Fazio}}]{Muller2026}%
  \BibitemOpen
  \bibfield  {author} {\bibinfo {author} {\bibfnamefont {T.~M.}\ \bibnamefont {M\"{u}ller}}, \bibinfo {author} {\bibfnamefont {S.}~\bibnamefont {Pappalardi}},\ and\ \bibinfo {author} {\bibfnamefont {R.}~\bibnamefont {Fazio}},\ }\bibfield  {title} {\bibinfo {title} {Quantum {M}pemba effect in chaotic systems with conservation laws},\ }\href {https://doi.org/10.1103/8nhn-1rs1} {\bibfield  {journal} {\bibinfo  {journal} {Phys. Rev. B}\ }\textbf {\bibinfo {volume} {114}},\ \bibinfo {pages} {L080301} (\bibinfo {year} {2026})}\BibitemShut {NoStop}%
\bibitem [{\citenamefont {Nava}\ and\ \citenamefont {Fabrizio}(2019)}]{Nava_Fabrizio_PRB2019_Mpemba}%
  \BibitemOpen
  \bibfield  {author} {\bibinfo {author} {\bibfnamefont {A.}~\bibnamefont {Nava}}\ and\ \bibinfo {author} {\bibfnamefont {M.}~\bibnamefont {Fabrizio}},\ }\bibfield  {title} {\bibinfo {title} {{Lindblad dissipative dynamics in the presence of phase coexistence}},\ }\href {https://doi.org/10.1103/PhysRevB.100.125102} {\bibfield  {journal} {\bibinfo  {journal} {Phys. Rev. B}\ }\textbf {\bibinfo {volume} {100}},\ \bibinfo {pages} {125102} (\bibinfo {year} {2019})}\BibitemShut {NoStop}%
\bibitem [{\citenamefont {Carollo}\ \emph {et~al.}(2021)\citenamefont {Carollo}, \citenamefont {Lasanta},\ and\ \citenamefont {Lesanovsky}}]{Carollo_Lesanovsky_PRL2021_Mpemba}%
  \BibitemOpen
  \bibfield  {author} {\bibinfo {author} {\bibfnamefont {F.}~\bibnamefont {Carollo}}, \bibinfo {author} {\bibfnamefont {A.}~\bibnamefont {Lasanta}},\ and\ \bibinfo {author} {\bibfnamefont {I.}~\bibnamefont {Lesanovsky}},\ }\bibfield  {title} {\bibinfo {title} {Exponentially accelerated approach to stationarity in {M}arkovian open quantum systems through the {M}pemba effect},\ }\href {https://doi.org/10.1103/PhysRevLett.127.060401} {\bibfield  {journal} {\bibinfo  {journal} {Phys. Rev. Lett.}\ }\textbf {\bibinfo {volume} {127}},\ \bibinfo {pages} {060401} (\bibinfo {year} {2021})}\BibitemShut {NoStop}%
\bibitem [{\citenamefont {Moroder}\ \emph {et~al.}(2024)\citenamefont {Moroder}, \citenamefont {Culhane}, \citenamefont {Zawadzki},\ and\ \citenamefont {Goold}}]{Moroder2024}%
  \BibitemOpen
  \bibfield  {author} {\bibinfo {author} {\bibfnamefont {M.}~\bibnamefont {Moroder}}, \bibinfo {author} {\bibfnamefont {O.}~\bibnamefont {Culhane}}, \bibinfo {author} {\bibfnamefont {K.}~\bibnamefont {Zawadzki}},\ and\ \bibinfo {author} {\bibfnamefont {J.}~\bibnamefont {Goold}},\ }\bibfield  {title} {\bibinfo {title} {Thermodynamics of the quantum {M}pemba effect},\ }\href {https://doi.org/10.1103/physrevlett.133.140404} {\bibfield  {journal} {\bibinfo  {journal} {Phys. Rev. Lett.}\ }\textbf {\bibinfo {volume} {133}},\ \bibinfo {pages} {140404} (\bibinfo {year} {2024})}\BibitemShut {NoStop}%
\bibitem [{\citenamefont {Ivander}\ \emph {et~al.}(2023)\citenamefont {Ivander}, \citenamefont {Anto-Sztrikacs},\ and\ \citenamefont {Segal}}]{Ivander_Segal_PRE2023_Mpemba}%
  \BibitemOpen
  \bibfield  {author} {\bibinfo {author} {\bibfnamefont {F.}~\bibnamefont {Ivander}}, \bibinfo {author} {\bibfnamefont {N.}~\bibnamefont {Anto-Sztrikacs}},\ and\ \bibinfo {author} {\bibfnamefont {D.}~\bibnamefont {Segal}},\ }\bibfield  {title} {\bibinfo {title} {{Hyperacceleration of quantum thermalization dynamics by bypassing long-lived coherences: An analytical treatment}},\ }\href {https://doi.org/10.1103/PhysRevE.108.014130} {\bibfield  {journal} {\bibinfo  {journal} {Phys. Rev. E}\ }\textbf {\bibinfo {volume} {108}},\ \bibinfo {pages} {014130} (\bibinfo {year} {2023})}\BibitemShut {NoStop}%
\bibitem [{\citenamefont {Chatterjee}\ \emph {et~al.}(2023)\citenamefont {Chatterjee}, \citenamefont {Takada},\ and\ \citenamefont {Hayakawa}}]{Chatterjee_Hayakawa_PRL2023_Mpemba}%
  \BibitemOpen
  \bibfield  {author} {\bibinfo {author} {\bibfnamefont {A.~K.}\ \bibnamefont {Chatterjee}}, \bibinfo {author} {\bibfnamefont {S.}~\bibnamefont {Takada}},\ and\ \bibinfo {author} {\bibfnamefont {H.}~\bibnamefont {Hayakawa}},\ }\bibfield  {title} {\bibinfo {title} {Quantum {M}pemba effect in a quantum dot with reservoirs},\ }\href {https://doi.org/10.1103/PhysRevLett.131.080402} {\bibfield  {journal} {\bibinfo  {journal} {Phys. Rev. Lett.}\ }\textbf {\bibinfo {volume} {131}},\ \bibinfo {pages} {080402} (\bibinfo {year} {2023})}\BibitemShut {NoStop}%
\bibitem [{\citenamefont {Nava}\ and\ \citenamefont {Egger}(2024)}]{Nava_Egger_PRL2024_Mpemba}%
  \BibitemOpen
  \bibfield  {author} {\bibinfo {author} {\bibfnamefont {A.}~\bibnamefont {Nava}}\ and\ \bibinfo {author} {\bibfnamefont {R.}~\bibnamefont {Egger}},\ }\bibfield  {title} {\bibinfo {title} {Mpemba effects in open nonequilibrium quantum systems},\ }\href {https://doi.org/10.1103/PhysRevLett.133.136302} {\bibfield  {journal} {\bibinfo  {journal} {Phys. Rev. Lett.}\ }\textbf {\bibinfo {volume} {133}},\ \bibinfo {pages} {136302} (\bibinfo {year} {2024})}\BibitemShut {NoStop}%
\bibitem [{\citenamefont {Wang}\ and\ \citenamefont {Wang}(2024)}]{Wang2024_Mpemba}%
  \BibitemOpen
  \bibfield  {author} {\bibinfo {author} {\bibfnamefont {X.}~\bibnamefont {Wang}}\ and\ \bibinfo {author} {\bibfnamefont {J.}~\bibnamefont {Wang}},\ }\bibfield  {title} {\bibinfo {title} {{Mpemba effects in nonequilibrium open quantum systems}},\ }\href {https://doi.org/10.1103/PhysRevResearch.6.033330} {\bibfield  {journal} {\bibinfo  {journal} {Phys. Rev. Res.}\ }\textbf {\bibinfo {volume} {6}},\ \bibinfo {pages} {033330} (\bibinfo {year} {2024})}\BibitemShut {NoStop}%
\bibitem [{\citenamefont {Longhi}(2024)}]{Longhi_APLQ2024_Mpemba}%
  \BibitemOpen
  \bibfield  {author} {\bibinfo {author} {\bibfnamefont {S.}~\bibnamefont {Longhi}},\ }\bibfield  {title} {\bibinfo {title} {{Bosonic Mpemba effect with non-classical states of light}},\ }\href {https://doi.org/10.1063/5.0234457} {\bibfield  {journal} {\bibinfo  {journal} {APL Quantum}\ }\textbf {\bibinfo {volume} {1}},\ \bibinfo {pages} {046110} (\bibinfo {year} {2024})}\BibitemShut {NoStop}%
\bibitem [{\citenamefont {Furtado}\ and\ \citenamefont {Santos}(2025)}]{Furtado_Santos_AnnPhys2025_Mpemba}%
  \BibitemOpen
  \bibfield  {author} {\bibinfo {author} {\bibfnamefont {J.}~\bibnamefont {Furtado}}\ and\ \bibinfo {author} {\bibfnamefont {A.~C.}\ \bibnamefont {Santos}},\ }\bibfield  {title} {\bibinfo {title} {{Enhanced quantum Mpemba effect with squeezed thermal reservoirs}},\ }\href {https://doi.org/https://doi.org/10.1016/j.aop.2025.170135} {\bibfield  {journal} {\bibinfo  {journal} {Ann. Phys.}\ }\textbf {\bibinfo {volume} {480}},\ \bibinfo {pages} {170135} (\bibinfo {year} {2025})}\BibitemShut {NoStop}%
\bibitem [{\citenamefont {Strachan}\ \emph {et~al.}(2025)\citenamefont {Strachan}, \citenamefont {Purkayastha},\ and\ \citenamefont {Clark}}]{Strachan2025}%
  \BibitemOpen
  \bibfield  {author} {\bibinfo {author} {\bibfnamefont {D.~J.}\ \bibnamefont {Strachan}}, \bibinfo {author} {\bibfnamefont {A.}~\bibnamefont {Purkayastha}},\ and\ \bibinfo {author} {\bibfnamefont {S.~R.}\ \bibnamefont {Clark}},\ }\bibfield  {title} {\bibinfo {title} {Non-{M}arkovian quantum {M}pemba effect},\ }\href {https://doi.org/10.1103/physrevlett.134.220403} {\bibfield  {journal} {\bibinfo  {journal} {Phys. Rev. Lett.}\ }\textbf {\bibinfo {volume} {134}},\ \bibinfo {pages} {220403} (\bibinfo {year} {2025})}\BibitemShut {NoStop}%
\bibitem [{\citenamefont {Di~Giulio}\ \emph {et~al.}(2025)\citenamefont {Di~Giulio}, \citenamefont {Turkeshi},\ and\ \citenamefont {Murciano}}]{DiGiulio2025}%
  \BibitemOpen
  \bibfield  {author} {\bibinfo {author} {\bibfnamefont {G.}~\bibnamefont {Di~Giulio}}, \bibinfo {author} {\bibfnamefont {X.}~\bibnamefont {Turkeshi}},\ and\ \bibinfo {author} {\bibfnamefont {S.}~\bibnamefont {Murciano}},\ }\bibfield  {title} {\bibinfo {title} {Measurement-induced symmetry restoration and quantum {M}pemba effect},\ }\href {https://doi.org/10.3390/e27040407} {\bibfield  {journal} {\bibinfo  {journal} {Entropy}\ }\textbf {\bibinfo {volume} {27}},\ \bibinfo {pages} {407} (\bibinfo {year} {2025})}\BibitemShut {NoStop}%
\bibitem [{\citenamefont {Ganguly}\ and\ \citenamefont {Agarwalla}()}]{ganguly2026measurement}%
  \BibitemOpen
  \bibfield  {author} {\bibinfo {author} {\bibfnamefont {K.}~\bibnamefont {Ganguly}}\ and\ \bibinfo {author} {\bibfnamefont {B.~K.}\ \bibnamefont {Agarwalla}},\ }\href@noop {} {\bibinfo {title} {Measurement induced faster symmetry restoration in quantum trajectories}},\ \Eprint {https://arxiv.org/abs/arXiv:2601.18458} {arXiv:2601.18458} \BibitemShut {NoStop}%
\bibitem [{\citenamefont {Ul{\v{c}}akar}\ \emph {et~al.}()\citenamefont {Ul{\v{c}}akar}, \citenamefont {Sharipov}, \citenamefont {Lagnese},\ and\ \citenamefont {Lenar{\v{c}}i{\v{c}}}}]{ulvcakar2025conserved}%
  \BibitemOpen
  \bibfield  {author} {\bibinfo {author} {\bibfnamefont {I.}~\bibnamefont {Ul{\v{c}}akar}}, \bibinfo {author} {\bibfnamefont {R.}~\bibnamefont {Sharipov}}, \bibinfo {author} {\bibfnamefont {G.}~\bibnamefont {Lagnese}},\ and\ \bibinfo {author} {\bibfnamefont {Z.}~\bibnamefont {Lenar{\v{c}}i{\v{c}}}},\ }\href@noop {} {\bibinfo {title} {Conserved quantities enable the quantum {M}pemba effect in weakly open systems}},\ \Eprint {https://arxiv.org/abs/arXiv:2511.16739} {arXiv:2511.16739} \BibitemShut {NoStop}%
\bibitem [{\citenamefont {Zhao}\ and\ \citenamefont {Hou}(2026)}]{Zhao2026}%
  \BibitemOpen
  \bibfield  {author} {\bibinfo {author} {\bibfnamefont {M.}~\bibnamefont {Zhao}}\ and\ \bibinfo {author} {\bibfnamefont {Z.}~\bibnamefont {Hou}},\ }\bibfield  {title} {\bibinfo {title} {Noise-induced quantum {M}pemba effect},\ }\href {https://doi.org/10.1038/s42005-026-02645-0} {\bibfield  {journal} {\bibinfo  {journal} {Commun. Phys.}\ }\textbf {\bibinfo {volume} {9}},\ \bibinfo {pages} {192} (\bibinfo {year} {2026})}\BibitemShut {NoStop}%
\bibitem [{\citenamefont {Ali}\ \emph {et~al.}(2026)\citenamefont {Ali}, \citenamefont {Zad}, \citenamefont {Hussain}, \citenamefont {Al-Kuwari}, \citenamefont {Kuniyil}, \citenamefont {Rahim}, \citenamefont {Ja{\v{s}}{\v{c}}ur},\ and\ \citenamefont {Haddadi}}]{Ali2026}%
  \BibitemOpen
  \bibfield  {author} {\bibinfo {author} {\bibfnamefont {A.}~\bibnamefont {Ali}}, \bibinfo {author} {\bibfnamefont {H.~A.}\ \bibnamefont {Zad}}, \bibinfo {author} {\bibfnamefont {M.~I.}\ \bibnamefont {Hussain}}, \bibinfo {author} {\bibfnamefont {S.}~\bibnamefont {Al-Kuwari}}, \bibinfo {author} {\bibfnamefont {H.}~\bibnamefont {Kuniyil}}, \bibinfo {author} {\bibfnamefont {M.~T.}\ \bibnamefont {Rahim}}, \bibinfo {author} {\bibfnamefont {M.}~\bibnamefont {Ja{\v{s}}{\v{c}}ur}},\ and\ \bibinfo {author} {\bibfnamefont {S.}~\bibnamefont {Haddadi}},\ }\bibfield  {title} {\bibinfo {title} {Quantum {M}pemba effect in a four‐site {B}ose–{H}ubbard model},\ }\href {https://doi.org/10.1002/prop.70089} {\bibfield  {journal} {\bibinfo  {journal} {Fortschr. Phys.}\ }\textbf {\bibinfo {volume} {74}},\ \bibinfo {pages} {e70089} (\bibinfo {year} {2026})}\BibitemShut {NoStop}%
\bibitem [{\citenamefont {Wei}\ \emph {et~al.}(2026)\citenamefont {Wei}, \citenamefont {Xu}, \citenamefont {Jiang}, \citenamefont {Hu},\ and\ \citenamefont {Pan}}]{Wei2026}%
  \BibitemOpen
  \bibfield  {author} {\bibinfo {author} {\bibfnamefont {Z.}~\bibnamefont {Wei}}, \bibinfo {author} {\bibfnamefont {M.}~\bibnamefont {Xu}}, \bibinfo {author} {\bibfnamefont {X.-P.}\ \bibnamefont {Jiang}}, \bibinfo {author} {\bibfnamefont {H.}~\bibnamefont {Hu}},\ and\ \bibinfo {author} {\bibfnamefont {L.}~\bibnamefont {Pan}},\ }\bibfield  {title} {\bibinfo {title} {Quantum {M}pemba effect in dissipative spin chains at criticality},\ }\href {https://doi.org/10.1007/s11433-025-2878-4} {\bibfield  {journal} {\bibinfo  {journal} {Sci. China Phys. Mech. Astron.}\ }\textbf {\bibinfo {volume} {69}},\ \bibinfo {pages} {240315} (\bibinfo {year} {2026})}\BibitemShut {NoStop}%
\bibitem [{\citenamefont {Joshi}\ \emph {et~al.}(2024)\citenamefont {Joshi}, \citenamefont {Franke}, \citenamefont {Rath}, \citenamefont {Ares}, \citenamefont {Murciano}, \citenamefont {Kranzl}, \citenamefont {Blatt}, \citenamefont {Zoller}, \citenamefont {Vermersch}, \citenamefont {Calabrese}, \citenamefont {Roos},\ and\ \citenamefont {Joshi}}]{Joshi2024}%
  \BibitemOpen
  \bibfield  {author} {\bibinfo {author} {\bibfnamefont {L.~K.}\ \bibnamefont {Joshi}}, \bibinfo {author} {\bibfnamefont {J.}~\bibnamefont {Franke}}, \bibinfo {author} {\bibfnamefont {A.}~\bibnamefont {Rath}}, \bibinfo {author} {\bibfnamefont {F.}~\bibnamefont {Ares}}, \bibinfo {author} {\bibfnamefont {S.}~\bibnamefont {Murciano}}, \bibinfo {author} {\bibfnamefont {F.}~\bibnamefont {Kranzl}}, \bibinfo {author} {\bibfnamefont {R.}~\bibnamefont {Blatt}}, \bibinfo {author} {\bibfnamefont {P.}~\bibnamefont {Zoller}}, \bibinfo {author} {\bibfnamefont {B.}~\bibnamefont {Vermersch}}, \bibinfo {author} {\bibfnamefont {P.}~\bibnamefont {Calabrese}}, \bibinfo {author} {\bibfnamefont {C.~F.}\ \bibnamefont {Roos}},\ and\ \bibinfo {author} {\bibfnamefont {M.~K.}\ \bibnamefont {Joshi}},\ }\bibfield  {title} {\bibinfo {title} {Observing the quantum {M}pemba effect in quantum simulations},\ }\href {https://doi.org/10.1103/physrevlett.133.010402} {\bibfield  {journal} {\bibinfo  {journal} {Phys. Rev. Lett.}\ }\textbf
  {\bibinfo {volume} {133}},\ \bibinfo {pages} {010402} (\bibinfo {year} {2024})}\BibitemShut {NoStop}%
\bibitem [{\citenamefont {Aharony~Shapira}\ \emph {et~al.}(2024)\citenamefont {Aharony~Shapira}, \citenamefont {Shapira}, \citenamefont {Markov}, \citenamefont {Teza}, \citenamefont {Akerman}, \citenamefont {Raz},\ and\ \citenamefont {Ozeri}}]{AharonyShapira2024}%
  \BibitemOpen
  \bibfield  {author} {\bibinfo {author} {\bibfnamefont {S.}~\bibnamefont {Aharony~Shapira}}, \bibinfo {author} {\bibfnamefont {Y.}~\bibnamefont {Shapira}}, \bibinfo {author} {\bibfnamefont {J.}~\bibnamefont {Markov}}, \bibinfo {author} {\bibfnamefont {G.}~\bibnamefont {Teza}}, \bibinfo {author} {\bibfnamefont {N.}~\bibnamefont {Akerman}}, \bibinfo {author} {\bibfnamefont {O.}~\bibnamefont {Raz}},\ and\ \bibinfo {author} {\bibfnamefont {R.}~\bibnamefont {Ozeri}},\ }\bibfield  {title} {\bibinfo {title} {Inverse {M}pemba effect demonstrated on a single trapped ion qubit},\ }\href {https://doi.org/10.1103/physrevlett.133.010403} {\bibfield  {journal} {\bibinfo  {journal} {Phys. Rev. Lett.}\ }\textbf {\bibinfo {volume} {133}},\ \bibinfo {pages} {010403} (\bibinfo {year} {2024})}\BibitemShut {NoStop}%
\bibitem [{\citenamefont {Zhang}\ \emph {et~al.}(2025{\natexlab{b}})\citenamefont {Zhang}, \citenamefont {Xia}, \citenamefont {Wu}, \citenamefont {Chen}, \citenamefont {Zhang}, \citenamefont {Xie}, \citenamefont {Su}, \citenamefont {Wu}, \citenamefont {Qiu}, \citenamefont {Chen}, \citenamefont {Li}, \citenamefont {Jing},\ and\ \citenamefont {Zhou}}]{Zhang2025}%
  \BibitemOpen
  \bibfield  {author} {\bibinfo {author} {\bibfnamefont {J.}~\bibnamefont {Zhang}}, \bibinfo {author} {\bibfnamefont {G.}~\bibnamefont {Xia}}, \bibinfo {author} {\bibfnamefont {C.-W.}\ \bibnamefont {Wu}}, \bibinfo {author} {\bibfnamefont {T.}~\bibnamefont {Chen}}, \bibinfo {author} {\bibfnamefont {Q.}~\bibnamefont {Zhang}}, \bibinfo {author} {\bibfnamefont {Y.}~\bibnamefont {Xie}}, \bibinfo {author} {\bibfnamefont {W.-B.}\ \bibnamefont {Su}}, \bibinfo {author} {\bibfnamefont {W.}~\bibnamefont {Wu}}, \bibinfo {author} {\bibfnamefont {C.-W.}\ \bibnamefont {Qiu}}, \bibinfo {author} {\bibfnamefont {P.-X.}\ \bibnamefont {Chen}}, \bibinfo {author} {\bibfnamefont {W.}~\bibnamefont {Li}}, \bibinfo {author} {\bibfnamefont {H.}~\bibnamefont {Jing}},\ and\ \bibinfo {author} {\bibfnamefont {Y.-L.}\ \bibnamefont {Zhou}},\ }\bibfield  {title} {\bibinfo {title} {Observation of quantum strong {M}pemba effect},\ }\href {https://doi.org/10.1038/s41467-024-54303-0} {\bibfield  {journal} {\bibinfo  {journal} {Nat. Commun.}\
  }\textbf {\bibinfo {volume} {16}},\ \bibinfo {pages} {301} (\bibinfo {year} {2025}{\natexlab{b}})}\BibitemShut {NoStop}%
\bibitem [{\citenamefont {Xia}\ \emph {et~al.}()\citenamefont {Xia}, \citenamefont {Zheng}, \citenamefont {Huang}, \citenamefont {Wu}, \citenamefont {Xie}, \citenamefont {Chen}, \citenamefont {Wu}, \citenamefont {Li}, \citenamefont {Jing}, \citenamefont {Zhang}, \citenamefont {Zhou},\ and\ \citenamefont {Chen}}]{xia2026observation}%
  \BibitemOpen
  \bibfield  {author} {\bibinfo {author} {\bibfnamefont {G.}~\bibnamefont {Xia}}, \bibinfo {author} {\bibfnamefont {Y.-J.}\ \bibnamefont {Zheng}}, \bibinfo {author} {\bibfnamefont {J.}~\bibnamefont {Huang}}, \bibinfo {author} {\bibfnamefont {C.-W.}\ \bibnamefont {Wu}}, \bibinfo {author} {\bibfnamefont {Y.}~\bibnamefont {Xie}}, \bibinfo {author} {\bibfnamefont {T.}~\bibnamefont {Chen}}, \bibinfo {author} {\bibfnamefont {W.}~\bibnamefont {Wu}}, \bibinfo {author} {\bibfnamefont {W.}~\bibnamefont {Li}}, \bibinfo {author} {\bibfnamefont {H.}~\bibnamefont {Jing}}, \bibinfo {author} {\bibfnamefont {J.}~\bibnamefont {Zhang}}, \bibinfo {author} {\bibfnamefont {Y.-L.}\ \bibnamefont {Zhou}},\ and\ \bibinfo {author} {\bibfnamefont {P.-X.}\ \bibnamefont {Chen}},\ }\href@noop {} {\bibinfo {title} {Observation of quantum multi-{M}pemba effect in a trapped-ion system}},\ \Eprint {https://arxiv.org/abs/arXiv:2604.21320} {arXiv:2604.21320} \BibitemShut {NoStop}%
\bibitem [{\citenamefont {Chatterjee}\ \emph {et~al.}(2025)\citenamefont {Chatterjee}, \citenamefont {Khan}, \citenamefont {Jain},\ and\ \citenamefont {Mahesh}}]{chatterjee2025_NMRMpemba}%
  \BibitemOpen
  \bibfield  {author} {\bibinfo {author} {\bibfnamefont {A.}~\bibnamefont {Chatterjee}}, \bibinfo {author} {\bibfnamefont {S.}~\bibnamefont {Khan}}, \bibinfo {author} {\bibfnamefont {S.}~\bibnamefont {Jain}},\ and\ \bibinfo {author} {\bibfnamefont {T.~S.}\ \bibnamefont {Mahesh}},\ }\href {https://arxiv.org/abs/2509.13451} {\bibinfo {title} {Direct experimental observation of quantum mpemba effect without bath engineering}} (\bibinfo {year} {2025}),\ \Eprint {https://arxiv.org/abs/2509.13451} {arXiv:2509.13451 [quant-ph]} \BibitemShut {NoStop}%
\bibitem [{\citenamefont {Xu}\ \emph {et~al.}(2026)\citenamefont {Xu}, \citenamefont {Fang}, \citenamefont {Chen}, \citenamefont {Wang}, \citenamefont {Ge}, \citenamefont {Shi}, \citenamefont {Liu}, \citenamefont {Deng}, \citenamefont {Zhao}, \citenamefont {Liu}, \citenamefont {Li}, \citenamefont {Li}, \citenamefont {Wang}, \citenamefont {Liang}, \citenamefont {Feng}, \citenamefont {Guo}, \citenamefont {Gu}, \citenamefont {He}, \citenamefont {Liu}, \citenamefont {Mei}, \citenamefont {Xiao}, \citenamefont {Yan}, \citenamefont {Yu}, \citenamefont {Yuan}, \citenamefont {Zhang}, \citenamefont {Wang}, \citenamefont {Liu}, \citenamefont {Song}, \citenamefont {Tian}, \citenamefont {Zhang}, \citenamefont {Zhang}, \citenamefont {Huang}, \citenamefont {Xiang}, \citenamefont {Zheng}, \citenamefont {Xu},\ and\ \citenamefont {Fan}}]{Xu2026}%
  \BibitemOpen
  \bibfield  {author} {\bibinfo {author} {\bibfnamefont {Y.}~\bibnamefont {Xu}}, \bibinfo {author} {\bibfnamefont {C.-P.}\ \bibnamefont {Fang}}, \bibinfo {author} {\bibfnamefont {B.-J.}\ \bibnamefont {Chen}}, \bibinfo {author} {\bibfnamefont {M.-C.}\ \bibnamefont {Wang}}, \bibinfo {author} {\bibfnamefont {Z.-Y.}\ \bibnamefont {Ge}}, \bibinfo {author} {\bibfnamefont {Y.-H.}\ \bibnamefont {Shi}}, \bibinfo {author} {\bibfnamefont {Y.}~\bibnamefont {Liu}}, \bibinfo {author} {\bibfnamefont {C.-L.}\ \bibnamefont {Deng}}, \bibinfo {author} {\bibfnamefont {K.}~\bibnamefont {Zhao}}, \bibinfo {author} {\bibfnamefont {Z.-H.}\ \bibnamefont {Liu}}, \bibinfo {author} {\bibfnamefont {T.-M.}\ \bibnamefont {Li}}, \bibinfo {author} {\bibfnamefont {H.}~\bibnamefont {Li}}, \bibinfo {author} {\bibfnamefont {Z.}~\bibnamefont {Wang}}, \bibinfo {author} {\bibfnamefont {G.-H.}\ \bibnamefont {Liang}}, \bibinfo {author} {\bibfnamefont {D.}~\bibnamefont {Feng}}, \bibinfo {author} {\bibfnamefont {X.-Y.}\ \bibnamefont {Guo}}, \bibinfo
  {author} {\bibfnamefont {X.-Y.}\ \bibnamefont {Gu}}, \bibinfo {author} {\bibfnamefont {Y.}~\bibnamefont {He}}, \bibinfo {author} {\bibfnamefont {H.-T.}\ \bibnamefont {Liu}}, \bibinfo {author} {\bibfnamefont {Z.-Y.}\ \bibnamefont {Mei}}, \bibinfo {author} {\bibfnamefont {Y.}~\bibnamefont {Xiao}}, \bibinfo {author} {\bibfnamefont {Y.}~\bibnamefont {Yan}}, \bibinfo {author} {\bibfnamefont {Y.-H.}\ \bibnamefont {Yu}}, \bibinfo {author} {\bibfnamefont {W.-P.}\ \bibnamefont {Yuan}}, \bibinfo {author} {\bibfnamefont {J.-C.}\ \bibnamefont {Zhang}}, \bibinfo {author} {\bibfnamefont {Z.-A.}\ \bibnamefont {Wang}}, \bibinfo {author} {\bibfnamefont {G.}~\bibnamefont {Liu}}, \bibinfo {author} {\bibfnamefont {X.}~\bibnamefont {Song}}, \bibinfo {author} {\bibfnamefont {Y.}~\bibnamefont {Tian}}, \bibinfo {author} {\bibfnamefont {Y.-R.}\ \bibnamefont {Zhang}}, \bibinfo {author} {\bibfnamefont {S.-X.}\ \bibnamefont {Zhang}}, \bibinfo {author} {\bibfnamefont {K.}~\bibnamefont {Huang}}, \bibinfo {author} {\bibfnamefont
  {Z.}~\bibnamefont {Xiang}}, \bibinfo {author} {\bibfnamefont {D.}~\bibnamefont {Zheng}}, \bibinfo {author} {\bibfnamefont {K.}~\bibnamefont {Xu}},\ and\ \bibinfo {author} {\bibfnamefont {H.}~\bibnamefont {Fan}},\ }\bibfield  {title} {\bibinfo {title} {Observation and modulation of the quantum {M}pemba effect on a superconducting quantum processor},\ }\href {https://doi.org/10.1103/951q-j8kq} {\bibfield  {journal} {\bibinfo  {journal} {Phys. Rev. Lett.}\ }\textbf {\bibinfo {volume} {137}},\ \bibinfo {pages} {010402} (\bibinfo {year} {2026})}\BibitemShut {NoStop}%
\bibitem [{\citenamefont {Summer}\ \emph {et~al.}(2026)\citenamefont {Summer}, \citenamefont {Moroder}, \citenamefont {Bettmann}, \citenamefont {Turkeshi}, \citenamefont {Marvian},\ and\ \citenamefont {Goold}}]{Summer2026}%
  \BibitemOpen
  \bibfield  {author} {\bibinfo {author} {\bibfnamefont {A.}~\bibnamefont {Summer}}, \bibinfo {author} {\bibfnamefont {M.}~\bibnamefont {Moroder}}, \bibinfo {author} {\bibfnamefont {L.~P.}\ \bibnamefont {Bettmann}}, \bibinfo {author} {\bibfnamefont {X.}~\bibnamefont {Turkeshi}}, \bibinfo {author} {\bibfnamefont {I.}~\bibnamefont {Marvian}},\ and\ \bibinfo {author} {\bibfnamefont {J.}~\bibnamefont {Goold}},\ }\bibfield  {title} {\bibinfo {title} {Resource-theoretical unification of {M}pemba effects: Classical and quantum},\ }\href {https://doi.org/10.1103/rbt4-psfd} {\bibfield  {journal} {\bibinfo  {journal} {Phys. Rev. X}\ }\textbf {\bibinfo {volume} {16}},\ \bibinfo {pages} {011065} (\bibinfo {year} {2026})}\BibitemShut {NoStop}%
\bibitem [{\citenamefont {Yamashika}\ and\ \citenamefont {Hamazaki}()}]{yamashika2026quantum}%
  \BibitemOpen
  \bibfield  {author} {\bibinfo {author} {\bibfnamefont {S.}~\bibnamefont {Yamashika}}\ and\ \bibinfo {author} {\bibfnamefont {R.}~\bibnamefont {Hamazaki}},\ }\href@noop {} {\bibinfo {title} {Quantum many-body {M}pemba effect through resonances}},\ \Eprint {https://arxiv.org/abs/arXiv:2603.11788} {arXiv:2603.11788} \BibitemShut {NoStop}%
\bibitem [{\citenamefont {Lu}\ and\ \citenamefont {Raz}(2017)}]{Lu_Raz_pnas2017_Mpemba}%
  \BibitemOpen
  \bibfield  {author} {\bibinfo {author} {\bibfnamefont {Z.}~\bibnamefont {Lu}}\ and\ \bibinfo {author} {\bibfnamefont {O.}~\bibnamefont {Raz}},\ }\bibfield  {title} {\bibinfo {title} {{Nonequilibrium thermodynamics of the Markovian Mpemba effect and its inverse}},\ }\href {https://doi.org/10.1073/pnas.1701264114} {\bibfield  {journal} {\bibinfo  {journal} {Proc. Natl Acad. Sci.}\ }\textbf {\bibinfo {volume} {114}},\ \bibinfo {pages} {5083} (\bibinfo {year} {2017})}\BibitemShut {NoStop}%
\bibitem [{\citenamefont {Li}\ \emph {et~al.}()\citenamefont {Li}, \citenamefont {Li},\ and\ \citenamefont {Yan}}]{li2025canonical}%
  \BibitemOpen
  \bibfield  {author} {\bibinfo {author} {\bibfnamefont {X.}~\bibnamefont {Li}}, \bibinfo {author} {\bibfnamefont {Y.}~\bibnamefont {Li}},\ and\ \bibinfo {author} {\bibfnamefont {Y.}~\bibnamefont {Yan}},\ }\href@noop {} {\bibinfo {title} {Canonical quantum {M}pemba effect in a dissipative qubit}},\ \Eprint {https://arxiv.org/abs/arXiv:2511.16996} {arXiv:2511.16996} \BibitemShut {NoStop}%
\bibitem [{\citenamefont {Hartmann}\ \emph {et~al.}(2004)\citenamefont {Hartmann}, \citenamefont {Keck}, \citenamefont {Korsch},\ and\ \citenamefont {Mossmann}}]{Hartmann2004}%
  \BibitemOpen
  \bibfield  {author} {\bibinfo {author} {\bibfnamefont {T.}~\bibnamefont {Hartmann}}, \bibinfo {author} {\bibfnamefont {F.}~\bibnamefont {Keck}}, \bibinfo {author} {\bibfnamefont {H.~J.}\ \bibnamefont {Korsch}},\ and\ \bibinfo {author} {\bibfnamefont {S.}~\bibnamefont {Mossmann}},\ }\bibfield  {title} {\bibinfo {title} {Dynamics of {B}loch oscillations},\ }\href {https://doi.org/10.1088/1367-2630/6/1/002} {\bibfield  {journal} {\bibinfo  {journal} {New J. Phys.}\ }\textbf {\bibinfo {volume} {6}},\ \bibinfo {pages} {2} (\bibinfo {year} {2004})}\BibitemShut {NoStop}%
\bibitem [{\citenamefont {Simon}\ \emph {et~al.}(2011)\citenamefont {Simon}, \citenamefont {Bakr}, \citenamefont {Ma}, \citenamefont {Tai}, \citenamefont {Preiss},\ and\ \citenamefont {Greiner}}]{Simon2011}%
  \BibitemOpen
  \bibfield  {author} {\bibinfo {author} {\bibfnamefont {J.}~\bibnamefont {Simon}}, \bibinfo {author} {\bibfnamefont {W.~S.}\ \bibnamefont {Bakr}}, \bibinfo {author} {\bibfnamefont {R.}~\bibnamefont {Ma}}, \bibinfo {author} {\bibfnamefont {M.~E.}\ \bibnamefont {Tai}}, \bibinfo {author} {\bibfnamefont {P.~M.}\ \bibnamefont {Preiss}},\ and\ \bibinfo {author} {\bibfnamefont {M.}~\bibnamefont {Greiner}},\ }\bibfield  {title} {\bibinfo {title} {Quantum simulation of antiferromagnetic spin chains in an optical lattice},\ }\href {https://doi.org/10.1038/nature09994} {\bibfield  {journal} {\bibinfo  {journal} {Nature}\ }\textbf {\bibinfo {volume} {472}},\ \bibinfo {pages} {307} (\bibinfo {year} {2011})}\BibitemShut {NoStop}%
\bibitem [{\citenamefont {Meinert}\ \emph {et~al.}(2014)\citenamefont {Meinert}, \citenamefont {Mark}, \citenamefont {Kirilov}, \citenamefont {Lauber}, \citenamefont {Weinmann}, \citenamefont {Gr\"obner},\ and\ \citenamefont {N\"agerl}}]{Meinert_Naegerl_PRL2014_BlochOscilExpt}%
  \BibitemOpen
  \bibfield  {author} {\bibinfo {author} {\bibfnamefont {F.}~\bibnamefont {Meinert}}, \bibinfo {author} {\bibfnamefont {M.~J.}\ \bibnamefont {Mark}}, \bibinfo {author} {\bibfnamefont {E.}~\bibnamefont {Kirilov}}, \bibinfo {author} {\bibfnamefont {K.}~\bibnamefont {Lauber}}, \bibinfo {author} {\bibfnamefont {P.}~\bibnamefont {Weinmann}}, \bibinfo {author} {\bibfnamefont {M.}~\bibnamefont {Gr\"obner}},\ and\ \bibinfo {author} {\bibfnamefont {H.-C.}\ \bibnamefont {N\"agerl}},\ }\bibfield  {title} {\bibinfo {title} {{Interaction-Induced Quantum Phase Revivals and Evidence for the Transition to the Quantum Chaotic Regime in 1D Atomic Bloch Oscillations}},\ }\href {https://doi.org/10.1103/PhysRevLett.112.193003} {\bibfield  {journal} {\bibinfo  {journal} {Phys. Rev. Lett.}\ }\textbf {\bibinfo {volume} {112}},\ \bibinfo {pages} {193003} (\bibinfo {year} {2014})}\BibitemShut {NoStop}%
\bibitem [{\citenamefont {Preiss}\ \emph {et~al.}(2015)\citenamefont {Preiss}, \citenamefont {Ma}, \citenamefont {Tai}, \citenamefont {Lukin}, \citenamefont {Rispoli}, \citenamefont {Zupancic}, \citenamefont {Lahini}, \citenamefont {Islam},\ and\ \citenamefont {Greiner}}]{Preiss_Greiner_Science2015_quantumwalks}%
  \BibitemOpen
  \bibfield  {author} {\bibinfo {author} {\bibfnamefont {P.~M.}\ \bibnamefont {Preiss}}, \bibinfo {author} {\bibfnamefont {R.}~\bibnamefont {Ma}}, \bibinfo {author} {\bibfnamefont {M.~E.}\ \bibnamefont {Tai}}, \bibinfo {author} {\bibfnamefont {A.}~\bibnamefont {Lukin}}, \bibinfo {author} {\bibfnamefont {M.}~\bibnamefont {Rispoli}}, \bibinfo {author} {\bibfnamefont {P.}~\bibnamefont {Zupancic}}, \bibinfo {author} {\bibfnamefont {Y.}~\bibnamefont {Lahini}}, \bibinfo {author} {\bibfnamefont {R.}~\bibnamefont {Islam}},\ and\ \bibinfo {author} {\bibfnamefont {M.}~\bibnamefont {Greiner}},\ }\bibfield  {title} {\bibinfo {title} {Strongly correlated quantum walks in optical lattices},\ }\href {https://doi.org/10.1126/science.1260364} {\bibfield  {journal} {\bibinfo  {journal} {Science}\ }\textbf {\bibinfo {volume} {347}},\ \bibinfo {pages} {1229–1233} (\bibinfo {year} {2015})}\BibitemShut {NoStop}%
\bibitem [{\citenamefont {Adler}\ \emph {et~al.}(2024)\citenamefont {Adler}, \citenamefont {Wei}, \citenamefont {Will}, \citenamefont {Srakaew}, \citenamefont {Agrawal}, \citenamefont {Weckesser}, \citenamefont {Moessner}, \citenamefont {Pollmann}, \citenamefont {Bloch},\ and\ \citenamefont {Zeiher}}]{Adler2024}%
  \BibitemOpen
  \bibfield  {author} {\bibinfo {author} {\bibfnamefont {D.}~\bibnamefont {Adler}}, \bibinfo {author} {\bibfnamefont {D.}~\bibnamefont {Wei}}, \bibinfo {author} {\bibfnamefont {M.}~\bibnamefont {Will}}, \bibinfo {author} {\bibfnamefont {K.}~\bibnamefont {Srakaew}}, \bibinfo {author} {\bibfnamefont {S.}~\bibnamefont {Agrawal}}, \bibinfo {author} {\bibfnamefont {P.}~\bibnamefont {Weckesser}}, \bibinfo {author} {\bibfnamefont {R.}~\bibnamefont {Moessner}}, \bibinfo {author} {\bibfnamefont {F.}~\bibnamefont {Pollmann}}, \bibinfo {author} {\bibfnamefont {I.}~\bibnamefont {Bloch}},\ and\ \bibinfo {author} {\bibfnamefont {J.}~\bibnamefont {Zeiher}},\ }\bibfield  {title} {\bibinfo {title} {Observation of {H}ilbert space fragmentation and fractonic excitations in {2D}},\ }\href {https://doi.org/10.1038/s41586-024-08188-0} {\bibfield  {journal} {\bibinfo  {journal} {Nature}\ }\textbf {\bibinfo {volume} {636}},\ \bibinfo {pages} {80} (\bibinfo {year} {2024})}\BibitemShut {NoStop}%
\bibitem [{\citenamefont {Haga}\ \emph {et~al.}(2021)\citenamefont {Haga}, \citenamefont {Nakagawa}, \citenamefont {Hamazaki},\ and\ \citenamefont {Ueda}}]{Haga_Ueda_PRL2021_LiouvillianSkinEffect}%
  \BibitemOpen
  \bibfield  {author} {\bibinfo {author} {\bibfnamefont {T.}~\bibnamefont {Haga}}, \bibinfo {author} {\bibfnamefont {M.}~\bibnamefont {Nakagawa}}, \bibinfo {author} {\bibfnamefont {R.}~\bibnamefont {Hamazaki}},\ and\ \bibinfo {author} {\bibfnamefont {M.}~\bibnamefont {Ueda}},\ }\bibfield  {title} {\bibinfo {title} {Liouvillian skin effect: Slowing down of relaxation processes without gap closing},\ }\href {https://doi.org/10.1103/PhysRevLett.127.070402} {\bibfield  {journal} {\bibinfo  {journal} {Phys. Rev. Lett.}\ }\textbf {\bibinfo {volume} {127}},\ \bibinfo {pages} {070402} (\bibinfo {year} {2021})}\BibitemShut {NoStop}%
\bibitem [{\citenamefont {Sharma}\ and\ \citenamefont {Mueller}(2021)}]{Sharma2021}%
  \BibitemOpen
  \bibfield  {author} {\bibinfo {author} {\bibfnamefont {V.}~\bibnamefont {Sharma}}\ and\ \bibinfo {author} {\bibfnamefont {E.~J.}\ \bibnamefont {Mueller}},\ }\bibfield  {title} {\bibinfo {title} {Driven-dissipative control of cold atoms in tilted optical lattices},\ }\href {https://doi.org/10.1103/physreva.103.043322} {\bibfield  {journal} {\bibinfo  {journal} {Phys. Rev. A}\ }\textbf {\bibinfo {volume} {103}},\ \bibinfo {pages} {043322} (\bibinfo {year} {2021})}\BibitemShut {NoStop}%
\bibitem [{\citenamefont {Garbe}\ \emph {et~al.}(2024)\citenamefont {Garbe}, \citenamefont {Minoguchi}, \citenamefont {Huber},\ and\ \citenamefont {Rabl}}]{Garbe2024}%
  \BibitemOpen
  \bibfield  {author} {\bibinfo {author} {\bibfnamefont {L.}~\bibnamefont {Garbe}}, \bibinfo {author} {\bibfnamefont {Y.}~\bibnamefont {Minoguchi}}, \bibinfo {author} {\bibfnamefont {J.}~\bibnamefont {Huber}},\ and\ \bibinfo {author} {\bibfnamefont {P.}~\bibnamefont {Rabl}},\ }\bibfield  {title} {\bibinfo {title} {The bosonic skin effect: Boundary condensation in asymmetric transport},\ }\href {https://doi.org/10.21468/scipostphys.16.1.029} {\bibfield  {journal} {\bibinfo  {journal} {SciPost Phys.}\ }\textbf {\bibinfo {volume} {16}},\ \bibinfo {pages} {029} (\bibinfo {year} {2024})}\BibitemShut {NoStop}%
\bibitem [{\citenamefont {Lindblad}(1976)}]{Lindblad_1976}%
  \BibitemOpen
  \bibfield  {author} {\bibinfo {author} {\bibfnamefont {G.}~\bibnamefont {Lindblad}},\ }\bibfield  {title} {\bibinfo {title} {{On the generators of quantum dynamical semigroups}},\ }\href {https://doi.org/10.1007/BF01608499} {\bibfield  {journal} {\bibinfo  {journal} {Commun. Math. Phys.}\ }\textbf {\bibinfo {volume} {48}},\ \bibinfo {pages} {119} (\bibinfo {year} {1976})}\BibitemShut {NoStop}%
\bibitem [{\citenamefont {Gorini}\ \emph {et~al.}(1976)\citenamefont {Gorini}, \citenamefont {Kossakowski},\ and\ \citenamefont {Sudarshan}}]{GKS_1976}%
  \BibitemOpen
  \bibfield  {author} {\bibinfo {author} {\bibfnamefont {V.}~\bibnamefont {Gorini}}, \bibinfo {author} {\bibfnamefont {A.}~\bibnamefont {Kossakowski}},\ and\ \bibinfo {author} {\bibfnamefont {E.~C.~G.}\ \bibnamefont {Sudarshan}},\ }\bibfield  {title} {\bibinfo {title} {{Completely positive dynamical semigroups of N‐level systems}},\ }\href {https://doi.org/10.1063/1.522979} {\bibfield  {journal} {\bibinfo  {journal} {J. Math. Phys.}\ }\textbf {\bibinfo {volume} {17}},\ \bibinfo {pages} {821} (\bibinfo {year} {1976})}\BibitemShut {NoStop}%
\bibitem [{\citenamefont {Breuer}\ and\ \citenamefont {Petruccione}(2007)}]{BreuerBook}%
  \BibitemOpen
  \bibfield  {author} {\bibinfo {author} {\bibfnamefont {H.-P.}\ \bibnamefont {Breuer}}\ and\ \bibinfo {author} {\bibfnamefont {F.}~\bibnamefont {Petruccione}},\ }\href {https://doi.org/10.1093/acprof:oso/9780199213900.001.0001} {\emph {\bibinfo {title} {{The Theory of Open Quantum Systems}}}}\ (\bibinfo  {publisher} {Oxford University Press},\ \bibinfo {address} {Oxford},\ \bibinfo {year} {2007})\BibitemShut {NoStop}%
\bibitem [{\citenamefont {Wannier}(1962)}]{Wannier1962Localization}%
  \BibitemOpen
  \bibfield  {author} {\bibinfo {author} {\bibfnamefont {G.~H.}\ \bibnamefont {Wannier}},\ }\bibfield  {title} {\bibinfo {title} {Dynamics of band electrons in electric and magnetic fields},\ }\href {https://doi.org/10.1103/RevModPhys.34.645} {\bibfield  {journal} {\bibinfo  {journal} {Rev. Mod. Phys.}\ }\textbf {\bibinfo {volume} {34}},\ \bibinfo {pages} {645} (\bibinfo {year} {1962})}\BibitemShut {NoStop}%
\bibitem [{\citenamefont {Zanardi}\ and\ \citenamefont {Campos~Venuti}(2014)}]{Zanardi2014_StrongDissPT}%
  \BibitemOpen
  \bibfield  {author} {\bibinfo {author} {\bibfnamefont {P.}~\bibnamefont {Zanardi}}\ and\ \bibinfo {author} {\bibfnamefont {L.}~\bibnamefont {Campos~Venuti}},\ }\bibfield  {title} {\bibinfo {title} {Coherent quantum dynamics in steady-state manifolds of strongly dissipative systems},\ }\href {https://doi.org/10.1103/PhysRevLett.113.240406} {\bibfield  {journal} {\bibinfo  {journal} {Phys. Rev. Lett.}\ }\textbf {\bibinfo {volume} {113}},\ \bibinfo {pages} {240406} (\bibinfo {year} {2014})}\BibitemShut {NoStop}%
\bibitem [{\citenamefont {Popkov}\ \emph {et~al.}(2018)\citenamefont {Popkov}, \citenamefont {Essink}, \citenamefont {Presilla},\ and\ \citenamefont {Sch\"utz}}]{Popkov2018_StrongDissPT}%
  \BibitemOpen
  \bibfield  {author} {\bibinfo {author} {\bibfnamefont {V.}~\bibnamefont {Popkov}}, \bibinfo {author} {\bibfnamefont {S.}~\bibnamefont {Essink}}, \bibinfo {author} {\bibfnamefont {C.}~\bibnamefont {Presilla}},\ and\ \bibinfo {author} {\bibfnamefont {G.}~\bibnamefont {Sch\"utz}},\ }\bibfield  {title} {\bibinfo {title} {Effective quantum zeno dynamics in dissipative quantum systems},\ }\href {https://doi.org/10.1103/PhysRevA.98.052110} {\bibfield  {journal} {\bibinfo  {journal} {Phys. Rev. A}\ }\textbf {\bibinfo {volume} {98}},\ \bibinfo {pages} {052110} (\bibinfo {year} {2018})}\BibitemShut {NoStop}%
\bibitem [{\citenamefont {Nielsen}\ and\ \citenamefont {Chuang}(2010)}]{NielsenChuang_2010}%
  \BibitemOpen
  \bibfield  {author} {\bibinfo {author} {\bibfnamefont {M.~A.}\ \bibnamefont {Nielsen}}\ and\ \bibinfo {author} {\bibfnamefont {I.~L.}\ \bibnamefont {Chuang}},\ }\href {https://doi.org/10.1017/CBO9780511976667} {\emph {\bibinfo {title} {Quantum Computation and Quantum Information: 10th Anniversary Edition}}}\ (\bibinfo  {publisher} {Cambridge University Press},\ \bibinfo {year} {2010})\BibitemShut {NoStop}%
\bibitem [{\citenamefont {van Kampen}(1992)}]{vanKampen1992_book}%
  \BibitemOpen
  \bibfield  {author} {\bibinfo {author} {\bibfnamefont {N.}~\bibnamefont {van Kampen}},\ }\href@noop {} {\emph {\bibinfo {title} {Stochastic Processes in Physics and Chemistry}}}\ (\bibinfo  {publisher} {Elsevier Science Publishers},\ \bibinfo {year} {1992})\BibitemShut {NoStop}%
\bibitem [{\citenamefont {Dias}\ \emph {et~al.}(2007)\citenamefont {Dias}, \citenamefont {Nascimento}, \citenamefont {Lyra},\ and\ \citenamefont {de~Moura}}]{DiasEtAl_PRB2007_2electronBlochOscil}%
  \BibitemOpen
  \bibfield  {author} {\bibinfo {author} {\bibfnamefont {W.~S.}\ \bibnamefont {Dias}}, \bibinfo {author} {\bibfnamefont {E.~M.}\ \bibnamefont {Nascimento}}, \bibinfo {author} {\bibfnamefont {M.~L.}\ \bibnamefont {Lyra}},\ and\ \bibinfo {author} {\bibfnamefont {F.~A. B.~F.}\ \bibnamefont {de~Moura}},\ }\bibfield  {title} {\bibinfo {title} {Frequency doubling of bloch oscillations for interacting electrons in a static electric field},\ }\href {https://doi.org/10.1103/PhysRevB.76.155124} {\bibfield  {journal} {\bibinfo  {journal} {Phys. Rev. B}\ }\textbf {\bibinfo {volume} {76}},\ \bibinfo {pages} {155124} (\bibinfo {year} {2007})}\BibitemShut {NoStop}%
\bibitem [{\citenamefont {Khomeriki}\ \emph {et~al.}(2010)\citenamefont {Khomeriki}, \citenamefont {Krimer}, \citenamefont {Haque},\ and\ \citenamefont {Flach}}]{Khomeriki_Krimer_Haque_Flach_PRA2010}%
  \BibitemOpen
  \bibfield  {author} {\bibinfo {author} {\bibfnamefont {R.}~\bibnamefont {Khomeriki}}, \bibinfo {author} {\bibfnamefont {D.~O.}\ \bibnamefont {Krimer}}, \bibinfo {author} {\bibfnamefont {M.}~\bibnamefont {Haque}},\ and\ \bibinfo {author} {\bibfnamefont {S.}~\bibnamefont {Flach}},\ }\bibfield  {title} {\bibinfo {title} {Interaction-induced fractional bloch and tunneling oscillations},\ }\href {https://doi.org/10.1103/PhysRevA.81.065601} {\bibfield  {journal} {\bibinfo  {journal} {Phys. Rev. A}\ }\textbf {\bibinfo {volume} {81}},\ \bibinfo {pages} {065601} (\bibinfo {year} {2010})}\BibitemShut {NoStop}%
\bibitem [{\citenamefont {Longhi}\ and\ \citenamefont {Della~Valle}(2012)}]{Longhi_DellaValle_PRB2012_2anyonBlochOscil}%
  \BibitemOpen
  \bibfield  {author} {\bibinfo {author} {\bibfnamefont {S.}~\bibnamefont {Longhi}}\ and\ \bibinfo {author} {\bibfnamefont {G.}~\bibnamefont {Della~Valle}},\ }\bibfield  {title} {\bibinfo {title} {Anyonic bloch oscillations},\ }\href {https://doi.org/10.1103/PhysRevB.85.165144} {\bibfield  {journal} {\bibinfo  {journal} {Phys. Rev. B}\ }\textbf {\bibinfo {volume} {85}},\ \bibinfo {pages} {165144} (\bibinfo {year} {2012})}\BibitemShut {NoStop}%
\bibitem [{\citenamefont {Wiater}\ \emph {et~al.}(2017)\citenamefont {Wiater}, \citenamefont {Sowi{\'n}ski},\ and\ \citenamefont {Zakrzewski}}]{Wiater_Zakrzewski_PRA2017_2bosons}%
  \BibitemOpen
  \bibfield  {author} {\bibinfo {author} {\bibfnamefont {D.}~\bibnamefont {Wiater}}, \bibinfo {author} {\bibfnamefont {T.}~\bibnamefont {Sowi{\'n}ski}},\ and\ \bibinfo {author} {\bibfnamefont {J.}~\bibnamefont {Zakrzewski}},\ }\bibfield  {title} {\bibinfo {title} {Two bosonic quantum walkers in one-dimensional optical lattices},\ }\href {https://doi.org/10.1103/PhysRevA.96.043629} {\bibfield  {journal} {\bibinfo  {journal} {Phys. Rev. A}\ }\textbf {\bibinfo {volume} {96}},\ \bibinfo {pages} {043629} (\bibinfo {year} {2017})}\BibitemShut {NoStop}%
\bibitem [{\citenamefont {Ribeiro}\ \emph {et~al.}(2020)\citenamefont {Ribeiro}, \citenamefont {Lazarides},\ and\ \citenamefont {Haque}}]{RibeiroLazaridesHaque_PRL2020}%
  \BibitemOpen
  \bibfield  {author} {\bibinfo {author} {\bibfnamefont {P.}~\bibnamefont {Ribeiro}}, \bibinfo {author} {\bibfnamefont {A.}~\bibnamefont {Lazarides}},\ and\ \bibinfo {author} {\bibfnamefont {M.}~\bibnamefont {Haque}},\ }\bibfield  {title} {\bibinfo {title} {Many-body quantum dynamics of initially trapped systems due to a stark potential: Thermalization versus bloch oscillations},\ }\href {https://doi.org/10.1103/PhysRevLett.124.110603} {\bibfield  {journal} {\bibinfo  {journal} {Phys. Rev. Lett.}\ }\textbf {\bibinfo {volume} {124}},\ \bibinfo {pages} {110603} (\bibinfo {year} {2020})}\BibitemShut {NoStop}%
\bibitem [{\citenamefont {Sarkar}\ and\ \citenamefont {Sowi{\'n}ski}(2020)}]{Sarkar_Sowinski_PRA2020}%
  \BibitemOpen
  \bibfield  {author} {\bibinfo {author} {\bibfnamefont {S.}~\bibnamefont {Sarkar}}\ and\ \bibinfo {author} {\bibfnamefont {T.}~\bibnamefont {Sowi{\'n}ski}},\ }\bibfield  {title} {\bibinfo {title} {Correlations in few two-component quantum walkers on a tilted lattice},\ }\href {https://doi.org/10.1103/PhysRevA.102.043326} {\bibfield  {journal} {\bibinfo  {journal} {Phys. Rev. A}\ }\textbf {\bibinfo {volume} {102}},\ \bibinfo {pages} {043326} (\bibinfo {year} {2020})}\BibitemShut {NoStop}%
\bibitem [{\citenamefont {Zhang}\ \emph {et~al.}(2024)\citenamefont {Zhang}, \citenamefont {Jiang},\ and\ \citenamefont {Li}}]{Zhang_EtAl_PRB2024_2doublonBlochOscil}%
  \BibitemOpen
  \bibfield  {author} {\bibinfo {author} {\bibfnamefont {K.-L.}\ \bibnamefont {Zhang}}, \bibinfo {author} {\bibfnamefont {X.-D.}\ \bibnamefont {Jiang}},\ and\ \bibinfo {author} {\bibfnamefont {Y.-Y.}\ \bibnamefont {Li}},\ }\bibfield  {title} {\bibinfo {title} {Two-doublon bloch oscillations in the mass-imbalanced extended fermi-hubbard model},\ }\href {https://doi.org/10.1103/PhysRevB.110.184304} {\bibfield  {journal} {\bibinfo  {journal} {Phys. Rev. B}\ }\textbf {\bibinfo {volume} {110}},\ \bibinfo {pages} {184304} (\bibinfo {year} {2024})}\BibitemShut {NoStop}%
\bibitem [{\citenamefont {Ferrari}\ \emph {et~al.}(2006)\citenamefont {Ferrari}, \citenamefont {Poli}, \citenamefont {Sorrentino},\ and\ \citenamefont {Tino}}]{Ferrari2006}%
  \BibitemOpen
  \bibfield  {author} {\bibinfo {author} {\bibfnamefont {G.}~\bibnamefont {Ferrari}}, \bibinfo {author} {\bibfnamefont {N.}~\bibnamefont {Poli}}, \bibinfo {author} {\bibfnamefont {F.}~\bibnamefont {Sorrentino}},\ and\ \bibinfo {author} {\bibfnamefont {G.~M.}\ \bibnamefont {Tino}},\ }\bibfield  {title} {\bibinfo {title} {Long-lived {B}loch oscillations with bosonic {Sr} atoms and application to gravity measurement at the micrometer scale},\ }\href {https://doi.org/10.1103/physrevlett.97.060402} {\bibfield  {journal} {\bibinfo  {journal} {Phys. Rev. Lett.}\ }\textbf {\bibinfo {volume} {97}},\ \bibinfo {pages} {060402} (\bibinfo {year} {2006})}\BibitemShut {NoStop}%
\bibitem [{\citenamefont {Gustavsson}\ \emph {et~al.}(2008)\citenamefont {Gustavsson}, \citenamefont {Haller}, \citenamefont {Mark}, \citenamefont {Danzl}, \citenamefont {Rojas-Kopeinig},\ and\ \citenamefont {N\"{a}gerl}}]{Gustavsson2008}%
  \BibitemOpen
  \bibfield  {author} {\bibinfo {author} {\bibfnamefont {M.}~\bibnamefont {Gustavsson}}, \bibinfo {author} {\bibfnamefont {E.}~\bibnamefont {Haller}}, \bibinfo {author} {\bibfnamefont {M.~J.}\ \bibnamefont {Mark}}, \bibinfo {author} {\bibfnamefont {J.~G.}\ \bibnamefont {Danzl}}, \bibinfo {author} {\bibfnamefont {G.}~\bibnamefont {Rojas-Kopeinig}},\ and\ \bibinfo {author} {\bibfnamefont {H.-C.}\ \bibnamefont {N\"{a}gerl}},\ }\bibfield  {title} {\bibinfo {title} {Control of interaction-induced dephasing of {B}loch oscillations},\ }\href {https://doi.org/10.1103/physrevlett.100.080404} {\bibfield  {journal} {\bibinfo  {journal} {Phys. Rev. Lett.}\ }\textbf {\bibinfo {volume} {100}},\ \bibinfo {pages} {080404} (\bibinfo {year} {2008})}\BibitemShut {NoStop}%
\bibitem [{\citenamefont {Wiseman}\ and\ \citenamefont {Milburn}(2009)}]{wiseman2009book}%
  \BibitemOpen
  \bibfield  {author} {\bibinfo {author} {\bibfnamefont {H.~M.}\ \bibnamefont {Wiseman}}\ and\ \bibinfo {author} {\bibfnamefont {G.~J.}\ \bibnamefont {Milburn}},\ }\href {https://doi.org/10.1017/CBO9780511813948} {\emph {\bibinfo {title} {{Quantum measurement and control}}}}\ (\bibinfo  {publisher} {Cambridge University Press},\ \bibinfo {address} {Cambridge},\ \bibinfo {year} {2009})\BibitemShut {NoStop}%
\bibitem [{\citenamefont {Kato}(1995)}]{Kato_book}%
  \BibitemOpen
  \bibfield  {author} {\bibinfo {author} {\bibfnamefont {T.}~\bibnamefont {Kato}},\ }\href {https://doi.org/https://doi.org/10.1007/978-3-642-66282-9} {\emph {\bibinfo {title} {Perturbation Theory for Linear Operators}}},\ \bibinfo {edition} {2nd}\ ed.\ (\bibinfo  {publisher} {Springer Berlin, Heidelberg},\ \bibinfo {year} {1995})\BibitemShut {NoStop}%
\end{thebibliography}%

\twocolumngrid

\onecolumngrid
\vspace{1em}
\begin{center}
    \textbf{End Matter}
\end{center}
\vspace{1em}
\twocolumngrid

\appendix

{\em Stark localization} ---
In the infinite-size limit the Stark Hamiltonian 
\begin{align}
    H_\text{Stark} &= J\sum_{j=-\infty}^{\infty} \left(b_j^\dagger b_{j+1} + b_{j+1}^\dagger b_{j}\right) - \Delta \sum_{j=-\infty}^{\infty} j b_j^\dagger b_j
\end{align}
can be diagonalized using the mapping $d_m := \sum_{j} \mathcal{J}_{j-m}(2J/\Delta)b_j$, so that $H_\text{Stark} = -\Delta \sum_{m} m d_m^\dagger d_m$. Here $\mathcal{J}_{j-m}(2J/\Delta)$ is the order $j-m$ Bessel function of the first kind. For $J\ll \Delta$, this can be approximated as 
\begin{align}
    \mathcal{J}_{j-m}\left(\frac{2J}{\Delta}\right) &\sim \frac{s(j-m)}{\vert j-m \vert!} \left(\frac{J}{\Delta}\right)^{\vert j-m \vert},
\end{align}
with $s(j-m)=1$ for $j\geq{m}$ and $s(j-m)=(-1)^{j-m}$ for $j<{m}$. 
Hence, the eigenstates have the position distribution
\begin{align}
    \vert \langle 0|b_j d_m^\dagger |0\rangle \vert ^2 = \vert \psi_m(j)\vert^2 \approx \frac{1}{(\vert j-m\vert!)^2} \left(\frac{J}{\Delta} \right)^{2\vert j-m \vert}.
\end{align}
We find that these infinite-size expressions are also excellent approximations for the tails of the eigenfunctions of finite-sized systems, even for eigenstates that are localized on the boundary of the chain.

{\em Liouvillian perturbation theory} ---  Consider the bosonic Lindbladian with incoherent hopping 
\begin{align}\label{Appendix:hopping_Lindbladian}
    \mathcal{L} &= -i[H, \cdot] + \Gamma \cdot \Gamma^\dagger - \frac{1}{2} \{\Gamma^\dagger \Gamma, \cdot \}, \quad
    \Gamma = \sqrt{\gamma} b_1^\dagger b_2, 
\end{align}
where $H$ is the Hamiltonian (\ref{eq:Hamiltonian}) made up of the Stark Hamiltonian coupled to an additional isolated site. We determine the spectrum and eigenmodes of (\ref{Appendix:hopping_Lindbladian}) by performing perturbative expansions in the incoherent hopping rate $\gamma$ \cite{Kato_book}, decomposing the Lindbladian as $\mathcal{L} = \mathcal{L}_0 + \gamma \mathcal{L}_1$,
where $\mathcal{L}_0=-i[H, \cdot]$ is the unitary part and $\gamma \mathcal{L}_1$ is the dissipator. We denote the projector on a given eigenspace of $\mathcal{L}$ by $\mathcal{P}$, and the corresponding eigenvalue by $\lambda$. Up to first order in $\gamma$ the perturbative expansions for eigenprojectors and eigenvalues are \cite{Kato_book}
\begin{align}
    \mathcal{P} &\approx \mathcal{P}^{(0)} + \gamma \mathcal{P}^{(1)} = \mathcal{P}^{(0)} - \gamma\left(\mathcal{P}^{(0)} \mathcal{L}_1 S + S \mathcal{L}_1 \mathcal{P}^{(0)}\right), \\
    \lambda &\approx \lambda^{(0)} + \gamma \lambda^{(1)} = \lambda^{(0)} + \gamma \frac{1}{\text{dim} \mathcal{P}^{(0)}} \text{tr}\left[ \mathcal{L}_1 \mathcal{P}^{(0)}\right].
\end{align}
Here $\mathcal{P}^{(0)}$ and $\lambda^{(0)}$ are the corresponding eigenprojector and eigenvalue of $\mathcal{L}_0$, and $S=S(\lambda^{(0)})$ is the reduced resolvent of $\mathcal{L}_0$, which for a semisimple eigenvalue $\lambda^{(0)}$ can be written as $S(\lambda^{(0)}) = - \sum_{\mu \neq \lambda^{(0)}} \frac{\mathcal{P}^{(0)}_\mu}{\lambda^{(0)} - \mu}$, $\mu$ and $\mathcal{P}^{(0)}_\mu$ denoting the other eigenvalues of $\mathcal{L}_0$ and the corresponding eigenprojectors. We write the eigenvalues of $\mathcal{L}_0$ as
\begin{align}
    \lambda_{jk}^{(0)} = -i(E_j-E_k),
\end{align}
where $j,k \in \{1, \dots, L\}$ and $E_{j,k}$ are the eigenenergies of the Hamiltonian. For a finite system size the eigenenergies will not follow the Stark ladder exactly, and hence the eigenvalues $\lambda_{jk}^{(0)}, j \neq k$ will be non-degenerate. The corresponding right and left eigenmodes are related to the single-particle eigenstates $|\psi_j\rangle$,
\begin{align}
    r_{jk}^{(0)} = |\psi_j\rangle\langle\psi_k|=l_{jk}^{(0)} \ (j \neq k),
\end{align}
so that the zeroth order eigenprojector can be written as $\mathcal{P}^{(0)}_{jk} = \text{tr} \left[ (l_{jk}^{(0)})^\dagger \ \cdot \ \right] r_{jk}^{(0)} = \langle \psi_j|\cdot |\psi_k\rangle |\psi_j\rangle\langle\psi_k|$. The first order correction to the eigenvalue is then
\begin{align}
    \lambda_{jk}^{(1)} = \text{tr}\left[\mathcal{P}^{(0)}_{jk} \mathcal{L}_1\right] = - \frac{1}{2} \left(|\psi_j(2)|^2 + |\psi_k(2)|^2 \right).
\end{align}
In the degenerate sector $j=k$ with $\lambda_{jj}^{(0)}=0$ the proper basis for the perturbative expansion is determined by diagonalizing the effective Lindbladian $\mathcal{P}^{(0)}\mathcal{L}\mathcal{P}^{(0)}$ on this subspace. Its matrix elements in the basis spanned by the Hamiltonian eigenstates are
\begin{align}
    (\mathcal{P}^{(0)} \mathcal{L} \mathcal{P}^{(0)})_{jk} &= \text{tr}\left[|\psi_j\rangle\langle\psi_j|\mathcal{P}^{(0)} \mathcal{L}_1 \mathcal{P}^{(0)}(|\psi_k\rangle\langle\psi_k|)\right] \nonumber \\
    &= \langle \psi_j|\mathcal{L}_1(|\psi_k\rangle\langle\psi_k|)|\psi_j\rangle \nonumber \\
    &= \delta_{j1} |\psi_k(2)|^2 - \delta_{jk} |\psi_k(2)|^2. 
\end{align}
Diagonalizing this matrix we obtain the right and left eigenmodes to zero'th order
\begin{align}
    r_{jj}^{(0)} &= 
    \begin{cases}
        |\psi_1\rangle\langle\psi_1|, &j=1 \\
        |\psi_j\rangle\langle\psi_j|-|\psi_1\rangle\langle\psi_1|, &j \geq 2
    \end{cases}, \\ 
    l_{jj}^{(0)} &= 
    \begin{cases}
        \mathbbm{1}, &j=1 \\
        |\psi_j\rangle\langle\psi_j|, &j\geq2
    \end{cases},
\end{align}
and the first order contribution to the eigenvalues
\begin{align}
    \lambda_{jj}^{(1)} &= -|\psi_j(2)|^2. 
\end{align}
Hence, to leading order in $\gamma$ the pure eigenstates of the Hamiltonian form the eigenmodes of the Lindbladian, and their decay rates are given by the weight their wavefunction has on the dissipative site. 
This lowest-order approximation works well upto moderate $\gamma$ because 
higher order corrections contain increasing powers of $\psi_j(2)$, so that localization suppresses these terms.

For $\gamma \gg \Delta$ a similar perturbative expansion can be performed. As detailed in \cite{Zanardi2014_StrongDissPT, Popkov2018_StrongDissPT} the strongly dissipative site 2 can be integrated out from the dynamics on long timescales, yielding an effective master equation with effective hopping $\Gamma_\text{eff} = \sqrt{\gamma_\text{eff}} \, b_1^\dagger b_3$ with effective hopping strength $\gamma_\text{eff} = \frac{4J^2}{\gamma}$. Since $\gamma$ is assumed to be large, the problem can be tackled again with a perturbative expansion in $\gamma_\text{eff}$, which gives completely analogous results to the weak-$\gamma$ case if one replaces $\gamma \to \gamma_\text{eff}$ and $\psi_j(2) \to \psi_j(3)$ in the formulas above. 

\begin{figure}
\resizebox{\columnwidth}{!}{\includegraphics{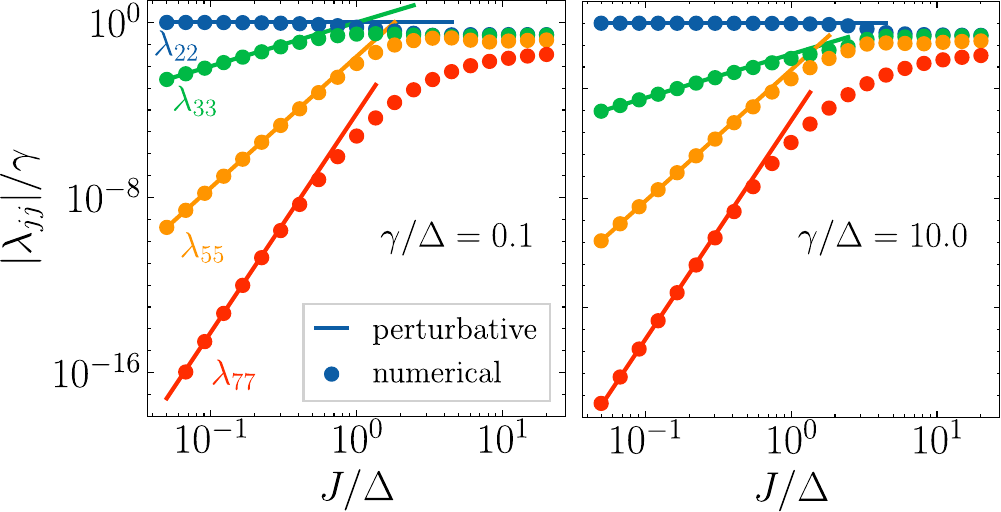}}
\caption{Real eigenvalues of the Lindbladian compared to perturbative expressions for weak (a) and strong (b) dissipation for $L=7$ and $\varepsilon/\Delta = 1$. 
}
\label{fig:realvalsJ}
\end{figure}

Fig. \ref{fig:realvalsJ} shows that the perturbative estimates match the numerically exact results so long as we are in the localized regime, $J/\Delta \lesssim1$.

In the intermediate regime where neither the weak nor strong $\gamma$ expansions should be valid we approximate the real eigenvalues using a Padé interpolation between these cases. It is given by
\begin{align}\label{eq:Pade}
    \lambda_{jj} &\approx - \frac{|\psi_j(2)|^2 \gamma}{1 + \frac{|\psi_j(2)|^2}{|\psi_j(3)|^2}\gamma_\text{eff}^2} = - \frac{|\psi_j(2)|^2\gamma}{1+\frac{1}{4J^2} \frac{|\psi_j(2)|^2}{|\psi_j(3)|^2}\gamma^2},
\end{align}
so that $\lambda_{jj} \sim -|\psi_j(2)|^2\gamma \ (\gamma \to 0)$ and $\lambda_{jj} \sim -|\psi_j(3)|^2\frac{4J^2}{\gamma} \ (\gamma \to \infty)$. We find that even away from the weak/strong $\gamma$ limits this estimate gives excellent agreement with exact numerics, as can be seen in Fig. \ref{fig:energies_localized} (d).

{\em Unraveling and waiting time distribution} ---
The Lindblad master equation can be written as
\begin{align}
    \dot{\rho} &= \mathcal{L}(\rho) = -i \left(H_\text{nH}\rho - \rho H_\text{nH}^\dagger\right) + \Gamma \rho \Gamma^\dagger,
\end{align}
where $H_\text{nH} = H - \frac{i}{2}\Gamma^\dagger \Gamma = H - \frac{i}{2}\gamma n_2(n_1+1)$ is the non-Hermitian effective Hamiltonian. The time evolved density matrix can then be written as an ensemble average over pure trajectories evolving under $H_\text{nH}$ while sampling the jump term $\Gamma \cdot \Gamma^\dagger$ stochastically \cite{BreuerBook}.
The cumulative distribution of the jump time is given by
\begin{align}
    p\left[\vert \psi(0)\rangle, t\right] &= 1 - \vert\vert e^{-iH_\text{nH}t}|\psi(0)\rangle\vert\vert^2.
\end{align}
To approximate this distribution we write the exponential in the eigenbasis of $H_\text{nH}$, $e^{-iH_\text{nH}t} = \sum_j e^{-i\lambda_jt}|r_j\rangle\langle l_j|$, and expand both eigenvalues and (right and left) eigenvectors perturbatively in $\gamma$. The leading terms that are relevant at long timespans are those obtained by the limit $\gamma \to 0, \gamma t = \text{const.}$, hence corrections to the eigenvectors are not relevant and the $\gamma$ dependence only needs to be accounted for in the exponential. Hence the leading approxmation is $e^{-iH_\text{nH}t} \approx \sum_j e^{-i(E_j+\gamma \lambda_j^{(1)})t}|\psi_j\rangle\langle\psi_j|$, where $\lambda_j^{(1)} = \langle \psi_j|\frac{-i}{2}\Gamma^\dagger \Gamma|\psi_j\rangle = \frac{-i}{2} |\psi_j(2)|^2$, so that the time-evolved trajectory becomes
\begin{align}
    |\psi(t)\rangle &= e^{-iH_\text{nH}t}|\psi(0)\rangle \nonumber \\ &\approx \sum_j e^{-\frac{1}{2}|\psi_j(2)|^2\gamma t} e^{-iE_jt}\langle\psi_j|\psi(0)\rangle|\psi_j\rangle.
\end{align}
The jump time distribution is then
\begin{align}
    p\left[|\psi(0)\rangle, t\right] &\underbrace{\approx}_{\gamma \ll \Delta}
        1 - \sum_{j \geq 2} e^{-\gamma |\psi_j(2)|^2t} |\langle \psi_j|\psi(0)\rangle|^2 
\end{align}
for $\gamma \ll \Delta$. The analogous result in the limit $\gamma \gg \Delta$ is obtained by replacing $\gamma \to \gamma_\text{eff} = \frac{4J^2}{\gamma}$ and $\psi_j(2) \to \psi_j(3)$ in the equation,
\begin{align}
    p\left[|\psi(0)\rangle, t\right] \underbrace{\approx}_{\gamma \gg \Delta} 1 - \sum_{j \geq 3} e^{-\frac{4J^2}{\gamma}|\psi_j(3)|^2t}|\langle \psi_j|\psi(0)\rangle|^2 
\end{align}
Since a single particle relaxes to the steady state after a single jump event, one recovers the estimate for the lifetime of a localized initial state from Eqs.\ (\ref{eq:relaxation_times_small}, \ref{eq:relaxation_times_large}).

{\em Intuitive picture for the Liouvillian spectrum} --- The structure of the Liouvillian spectrum can be understood intuitively using the fact that the Hamiltonian eigenstates $|\psi_j\rangle$ are also approximate eigenmodes of the dissipative dynamics with decay rates $\Gamma_j \approx \gamma |\psi_j(2)|^2$. For an initial superposition $|\psi(0)\rangle = c_j|\psi_j\rangle + c_k |\psi_k\rangle$, $1<j<k$, the no-jump evolution under $H_\text{nH}$ is given by $|\psi(t)\rangle \approx c_j(t)|\psi_j\rangle + c_k(t)|\psi_k\rangle$ where $c_j(t) = e^{-iE_jt} e^{-\frac{1}{2}\Gamma_jt}$. Hence, the corresponding populations decay as 
\begin{align}
    |c_j(t)|^2 \sim e^{-\Gamma_jt},
\end{align}
whereas a coherence varies as 
\begin{align}
    c_j(t) c_k^*(t) \sim e^{-i(E_j-E_k)t} e^{-\frac{1}{2}(\Gamma_j+\Gamma_k)t} \approx e^{-i(k-j)\Delta t} e^{-\frac{1}{2}\Gamma_j t}. 
\end{align}

{\em Analog classical master equation} ---
For the classical analog, the particle occupation is described by a probability distribution $p_j(t)$ evolving under a Markovian master equation $\dot{\vec{p}}(t) = \mathcal{K} \vec{p}(t)$. The process allows for one-way transitions with rate $R$ from site 2 to site 1, and transitions with rate $C \ll R$ between adjacent sites not including site 1. This can be thought of as a thermal process with temperature $T \gg \Delta$ and $T \ll V_2 - V_1$. The Markov (Kolmogorov) matrix $\mathcal{K}$ is given by
\begin{align}
    \mathcal{K} &= 
    \begin{pmatrix}
        0 & R & & & & \\
        0 & -(R+C) & C & & & \\
        & C & -2C & C & & \\
        & & \ddots & \ddots &\ddots & \\
        & & & C & -2C & C \\
        & & & & C & -C
    \end{pmatrix}
\end{align}
Since this equation does not allow for transitions from site 1 to 2, the exact steady state is always be the state localized on the first site, $p_j(t\to \infty) = \delta_{j,1}$. Such a reverse transition may be included without changing the resulting dynamics substantially, however, so long as the corresponding transition rate is small compared to $R$.

\end{document}